\documentclass[preprint,12pt,sort&compress]{elsarticle}
\usepackage[utf8]{inputenc}
\usepackage{color}
\usepackage{graphicx}
\usepackage{placeins}
\usepackage[margin=1in]{geometry}
\usepackage{appendix}
\usepackage{amsmath, amssymb, bm}
\usepackage{multirow}
\usepackage{placeins}
\usepackage{xfrac}
\usepackage[colorlinks=true, linkcolor=blue, citecolor=blue]{hyperref}
\usepackage[nameinlink, capitalize]{cleveref}
\usepackage[font=footnotesize, labelfont=bf]{caption}
\usepackage{subcaption}
\usepackage[version=4]{mhchem}
\usepackage{lineno}
\usepackage{bm}
\usepackage{comment}
\usepackage[T1]{fontenc}
\usepackage{anyfontsize}

\usepackage{tablefootnote}
\usepackage{float}
\usepackage{multicol}
\usepackage{multirow}
\usepackage{tikz}
\usetikzlibrary{arrows.meta, positioning, fit}
\usepackage{nicefrac}

\journal{Journal of Nuclear Materials}

\begin{document}

\begin{frontmatter}

\title{A model of grain growth in UN integrating molecular dynamics, phase-field modeling, and uncertainty quantification}

\author[ncsu,unsw]{Mohamed AbdulHameed}
\author[ncsu,inl]{Fadel M. Nasr}
\author[ncsu,inl]{Wen Jiang}
\author[psu]{Mahmoud Yaseen}
\author[ncsu,inl]{Benjamin Beeler\corref{1}}
\cortext[1]{Corresponding author}
\ead{bwbeeler@ncsu.edu}
\address[ncsu]{Department of Nuclear Engineering, North Carolina State University, Raleigh, NC 27695, USA}
\address[inl]{Idaho National Laboratory, Idaho Falls, ID 83415, USA}
\address[psu]{Department of Nuclear Engineering, Pennsylvania State University, University Park, PA 16802, USA}
\address[unsw]{School of Mechanical and Manufacturing Engineering, UNSW Sydney, Sydney, NSW 2052, Australia}

\begin{abstract}

Grain growth kinetics and grain-boundary (GB) properties in uranium mononitride (UN) are investigated through an integrated multiscale framework combining molecular dynamics (MD), phase-field modeling, and surrogate-assisted uncertainty quantification. MD simulations yield GB energies for 27 symmetric tilt boundaries from 0--2000~K, which are consistent with available DFT values. The average GB energy is nearly temperature-independent below 1000~K and increases at higher temperatures. A mechanistic pore-drag model applied to the only available grain growth dataset for actinide nitrides yields a mobility reduction factor of $s \approx 0.93$--$0.99$, statistically indistinguishable from unity, confirming that pore drag is negligible under the experimental conditions. The intrinsic GB mobility is therefore extracted directly from the effective mobility, yielding $M_0 = 2.05\times10^{-15}$~m$^4$/(J$\cdot$s) and $Q_M = 0.89$~eV. Phase-field simulations conducted from 1500--2000~K confirm normal curvature-driven grain growth, with grain size distributions converging to the Hillert-like form. A surrogate-assisted global sensitivity analysis---combining principal component analysis, Gaussian process regression, and Sobol decomposition---reveals that the mobility prefactor $M_0$ dominates output variance at all times, followed by the activation energy $Q_M$, while the GB energy $\gamma$ contributes minimally. These results establish the first quantitative grain growth framework for UN and identify the reduction of uncertainty in $M_0$ and $Q_M$ as the highest-priority target for future experimental efforts.

\end{abstract}


\end{frontmatter}

\newpage

\section{Introduction}

Uranium mononitride (UN) stands out as a strong candidate for advanced nuclear fuels because of its exceptional characteristics: high fissile-material density, excellent thermal conductivity, broad compatibility with various cladding options, and the prospect of prolonged fuel residence times~\cite{Wallenius2020, Uno2020}. However, several factors continue to impede its widespread adoption. These include the intricate nature of its fabrication processes, the substantial expense associated with enriching $^{{15}}\mathrm{N}$, and a pronounced vulnerability to oxidation at elevated temperatures~\cite{Wallenius2020, Uno2020}. Although some studies attempted to give a mechanistic understanding of some interesting phenomena in UN, e.g., paramagnetism \cite{AIMD+DLM}, high-temperature specific heat \cite{AbdulHameed2024,FrenkelPairs}, dislocation motion \cite{AbdulHameed2024b}, thermal creep \cite{AbdulHameed2024c}, diffusivity under irradiation \cite{Cooper2023,Schneider2024}, and swelling \cite{Rizk2025,Abdulhameed2025Oygen}, many aspects of its microstructural response remain unexplored due to a scarcity of experimental studies and a limited understanding of the underlying mechanisms. In this context, the kinetics and mechanisms of grain growth in UN, a key microstructural process, have not yet been systematically explored.

In polycrystalline materials, grain boundaries (GBs) migrate to reduce total free energy, collectively leading to grain growth. This phenomenon is typically studied under isothermal conditions \cite{Ainscough1973}. GB migration may be driven by GB energy, elastic energy, or thermal gradients \cite{Tonks2014}, with curvature-driven grain growth being the dominant mechanism under typical conditions. Grain growth and GB mobility are inherently interconnected and have almost always been studied concurrently. Kaoumi \textit{et al.} \cite{Kaoumi2007,Kaoumi2008} proposed a model based on thermal-spike morphology to describe grain growth under irradiation in pure metals. Powers and Glaeser \cite{Powers1998} reviewed grain growth and GB mobility in ceramics, highlighting the effects of solute drag, pore-boundary interactions, liquid phases, and anisotropy on microstructural evolution. For nuclear fuels, French and Bai \cite{French2022} conducted an atomistic study to investigate GB mobility in $\gamma$-U. Additionally, there exist several experimental \cite{Ainscough1973}, atomistic \cite{Bai2015}, and mesoscale \cite{Tonks2014,Tonks2021} studies on grain growth and GB mobility for UO$_2$. Cheniour \textit{et al.} \cite{Cheniour2020} measured GB mobility in U$_3$Si$_2$ and conducted a phase-field model of its grain growth using the MOOSE framework \cite{MOOSE,PhaseField}. For UN, however, only the experimental study by Ronchi and Sari \cite{Ronchi1975} tried to quantify thermal grain growth in actinide nitrides, which was conducted on (U, Pu)N rather than pure UN. Johnson and Lopes \cite{Johnson2018} prepared UN by spark plasma sintering and observed that grain rotation and coalescence are the dominant mechanisms of grain growth in UN during sintering. However, no attempt was made to quantify this behavior.

An important parameter that is needed to perform mesoscale grain evolution simulations is the GB energy and its dependence on temperature and, possibly, misorientation. Molecular dynamics (MD) simulations allow us to study the static and dynamic properties of GBs at a moderate computational cost, whereas alternative computational methodologies, e.g., DFT and \textit{ab initio} MD, are limited by the requirements of large supercells and long simulation times, respectively, for the study of GB properties. Bicrystal models of tilt GBs have been utilized in MD to study the structure and energetics of GBs in Cu and Al \cite{Tschopp2007, Tschopp2007b}, Fe \cite{Shibuta2008}, and U$_3$Si$_2$ \cite{Beeler2019}, for example. Only the DFT study by Wang \textit{et al.} \cite{Wang2023} calculated the GB energy of UN. The calculation was done for only two symmetric tilt GBs, $\Sigma5(310)[001]$ and $\Sigma5(210)[001]$, at 0~K with no attempt to find the minimum energy structure of the GB using rigid body translations.

In this work, MD is used to calculate the GB energy for 27 symmetric tilt GBs in the temperature range of 0--2000~K. Then, a method is proposed to estimate the drag-free GB mobility of UN from the phenomenological grain growth formula. The estimated GB energy and mobility are utilized in a phase-field model of isothermal curvature-driven grain growth. Finally, a parametric study is performed to measure the sensitivity of the simulation output (average grain size) to uncertainties in the input parameters (GB energy and GB mobility).

\section{Methods}

\subsection{Grain growth formalism}

% Note that the effects of triple junctions and other sources of drag are not considered in this study. Instead, we focus solely on the curvature driving force of grain growth.

% Triple junctions (TJs) are a fundamental morphological feature of polycrystals, and are typically assumed to move with sufficient velocity to accommodate the migration of GBs \cite{Gottstein2000}. However, experimental studies \cite{Gottstein1999, Protasova2001} have shown that, under certain topological conditions, TJs can exert a drag force on GB movement, making TJ mobility the rate-limiting factor of grain growth rather than GB mobility. Additionally, solutes, porosity, and second-phase particles also introduce resistive forces that impede GB migration \cite{Tonks2021}. 

The velocity $v$ of a GB at a given location is described by \cite{Hillert1965, Tonks2021}:
\begin{equation}
v = \frac{dR}{dt} = M \left[ F - P \right]^+,
\end{equation}
where $R$ is the local GB radius of curvature, $M$ is the GB mobility, $F$ is the driving force per unit area, and $P$ is the resistive force per unit area. The functional operator $[x]^+$ equals $x$ when $x > 0$ and 0 otherwise \cite{Tonks2021}. The driving force due to GB curvature is given by \cite{Cheniour2020}:
\begin{equation}
F = \frac{\gamma}{R},
\end{equation}
where $\gamma$ is the GB energy. While GB properties are anisotropic, the overall grain growth behavior is often unaffected by this anisotropy \cite{Tonks2021}, justifying the use of average GB properties and grain sizes. Assuming a geometric relation between the average grain size $ D $ and the average curvature radius $\bar{R}$, given by $D = \alpha^{1/2} \bar{R}$ \cite{Cheniour2020}, the grain growth rate can be approximated as:
\begin{equation}
\frac{d D}{d t} \approx \alpha M \frac{\gamma}{D},
\label{Eq:Curve}
\end{equation}
where $\alpha = 1$ for three-dimensional (3D) growth and $\alpha = 1/2$ for two-dimensional (2D) growth \cite{Tonks2021}. $M$ and $\gamma$ are the GB mobility and GB energy as previously defined. Integrating under the assumption of constant $M$ and $\gamma$ gives:
\begin{equation}
D^2 - D_0^2 = 2 \alpha M \gamma t,
\label{Eq:ideal}
\end{equation}
where $D_0$ is the initial average grain size. In addition to the power law of \cref{Eq:ideal}, a key feature of isothermal curvature-driven grain growth is that, after reaching steady-state, the grain-size distribution exhibits time self-similarity, most commonly resembling a lognormal or Hillert distribution \cite{Hillert1965, Protasova2001}.

Although this ideal model holds for ultrapure materials close to their melting point \cite{Hu1970}, deviations are frequently observed experimentally due to resistive forces. To account for these complexities, the following semi-empirical model is used:
\begin{equation}
D^n - D_0^n = Kt,
\label{Eq:actual}
\end{equation}
where $K$ is the growth rate constant and $n$ is an empirically fitted exponent, typically between 2 and 4 \cite{Hu1970}, influenced by porosity, impurities, and residual stresses. Experimental data frequently suggest $n = 3$, especially in porous materials \cite{Kingery1965, Hu1970}. When $n = 2$, the model aligns with the ideal case, yielding:
\begin{equation}
K = 2 \alpha M \gamma.
\label{eq:Kn=2}
\end{equation}
For $n > 2$, $K$ becomes a function of additional factors, including temperature-gradient-driven forces, porosity-induced resistance, and other microstructural features \cite{Powers1998}.

\subsection{Grain-boundary energy}

All MD calculations performed in this work utilize the Large-scale Atomic/Molecular Massively Parallel Simulator (LAMMPS) software package \cite{Thompson2022} using a 1~fs time step and the Tseplyaev \cite{Tseplyaev2016} and Kocevski \cite{Kocevski2022II} interatomic potentials. Periodic boundary conditions (PBCs) are applied to all supercells. The Atomsk code \cite{Hirel2015} is used to generate bicrystals by the Voronoi tessellation method. The OVITO software package \cite{OVITO} is used for supercell visualization and analysis. 

The energy of the GB (per unit area) is a complex function of the five degrees of freedom of the GB and is often plotted against only one degree of freedom while the other four are kept fixed \cite{Cai2016}. Three degrees of freedom specify the orientation of one grain relative to another, and the other two degrees of freedom specify the orientation of the boundary relative to one of the grains. In this work, Atomsk \cite{Hirel2015} is used to generate bicrystals containing 27 symmetric tilt GBs in the center of the supercell with misorientation angles between 0$^\circ$ and 90$^\circ$ as detailed in \cref{Tab:GBPlanes}. The GB plane is the $xz$ plane and is normal to the $y$ direction, with a tilt axis of [001]. By construction and due to PBCs, another GB exists at $y = 0$, which is equivalent to $y = l_y$, with $l_y$ being the dimension of the supercell along the $y$-axis. Periodicity along the $x$-axis is maintained for a plane of $(ijk)$ by applying the following formulae:
\begin{equation}
p = \sqrt{i^2+j^2+k^2} \ a,
\end{equation}
\begin{equation}
l_x = 2p \cdot \text{ceil}\left(\frac{8a}{p}\right),
\end{equation}
where $a$ is the lattice constant, and $p$ is the structural periodicity of the GB plane \cite{Abbaschian2025}, which is the distance between two consecutive atoms on the GB plane that are shared between the two grains. For all cases, $l_y$ is arbitrarily set to $l_x$, and $l_z$ is set to $8a$. After supercell construction, rigid-body translations of one grain relative to the other are performed \cite{Tschopp2007}. First, the lower grain (i.e., the grain between $y = 0$ and $y=l_y/2$) is displaced relative to the upper grain (i.e., the grain between $y = l_y/2$ and $y=l_y$) in the $x$-axis by distances between $p/8$ and $p$ with an increment of $p/8$ (8 configurations). Then, the lower grain is displaced by the same distances and the same increment in the $y$-axis (8 configurations). For displacements along the $z$-axis, only two configurations were used due to PBCs: one with no displacements, and another with a displacement of $a/2$ \cite{Beeler2019} (2 configurations). In total, rigid-body translations give 128 different configurations for each GB plane. Finally, an atom deletion iteration was performed for each of the 128 configurations, where atoms within distances between 0.2 \AA\ and 2 \AA, in increments of 0.2 \AA, are deleted (10 configurations). The aim of performing iterations of grain translations and atom deletion is to reach the true minimum energy structure of each GB plane. This method of finding the most stable GB structure is commonly called the $\gamma$-surface method \cite{Seki2023}. In total, for each GB plane, we have 1280 configurations, which are used for 0~K and finite-temperature calculations.

\begin{table}[ht]
\footnotesize
\centering
\caption{Grain-boundary planes studied in this work along with their misorientation angles around the [001] axis, labeled $\theta$, as well as their $\Sigma$ numbers \cite{Grimmer1974, Zhao1988, Wang2023}. All symmetric tilts GBs have inherent periodicity and are CSLs \cite{Riet2021}. However, only $\Sigma \leq$ 49 is shown.}
\begin{tabular}{ccc|ccc|ccc}
\hline
Plane & $\theta$ & $\Sigma$ & Plane & $\theta$ & $\Sigma$ & Plane & $\theta$ & $\Sigma$ \\
\hline
(910) & 12.680  & $41a$ & (830) & 41.112  &       & (320) & 67.380  &  \\
(810) & 14.250  &       & (520) & 43.603  & $29a$ & (750) & 71.075  &  \\
(710) & 16.260  & $25a$ & (730) & 46.397  &       & (430) & 73.740  &  \\
(610) & 18.925  & $37a$ & (940) & 47.925  &       & (970) & 75.750  &  \\
(510) & 22.620  & $13a$ & (210) & 53.130  & 5     & (540) & 77.320  &  \\
(920) & 25.058  &       & (950) & 58.109  &       & (650) & 79.611  &  \\
(410) & 28.072  & $17a$ & (740) & 59.490  &       & (760) & 81.203  &  \\
(720) & 31.891  &       & (530) & 61.928  &       & (870) & 82.372  &  \\
(310) & 36.870  & 5     & (850) & 64.011  &       & (980) & 83.267  &  \\ 
\hline
\end{tabular}
\label{Tab:GBPlanes}
\end{table}

GB energies are calculated at 0~K and finite temperatures using the formula \cite{Beeler2019}:
\begin{equation}
E_\text{GB} = \frac{E^* - N E_a}{A},
\label{Eq:GBE}
\end{equation}
where $E^*$ is the energy of the supercell containing the GB, $E_a$ is the average energy per atom in the perfect crystal (calculated as half of the cohesive energy; $E_a = E_c / 2$), $N$ is the total number of atoms in the bicrystal, and $A$ is the GB area calculated as $A = 2 l_x l_z$, where the factor of 2 in the GB area accounts for the two interfaces present in the system. Note that the bicrystals are guaranteed to be stoichiometric because they contain symmetric tilt GBs.

To get the GB energy at 0~K, the energy of each of the 1280 configurations is minimized with a relative energy tolerance of $10^{-12}$. A perfect $6 \times 6 \times 6$ UN crystal is also minimized using the same tolerance to get the cohesive energy at 0~K. To get the GB energy at finite temperatures, each of the 1280 configurations is cooled within the \textit{NPT} ensemble from 2500~K to the target temperature for 20 ps to allow the GB to reconfigure to the minimum energy structure. Then, the system is equilibrated at the target temperature for 100 ps, where the potential energy is averaged over the final 50 ps. To get the cohesive energies at finite temperatures, a perfect $15 \times 15 \times 15$ UN crystal is equilibrated at the target temperature for 100 ps, where the potential energy is also averaged over the final 50 ps.

\subsection{Phase-field modeling}
\label{Sec:PhaseField}

Grain growth simulations are conducted using the phase-field module \cite{PhaseField} of the Multiphysics Object Oriented Simulation Environment (MOOSE) framework \cite{MOOSE}. The phase-field model of grain growth implemented in MOOSE is based on the works by Chen and Yang \cite{Chen1994, Chen1995} and Moelans \textit{et al.} \cite{Moelans2008}. In this model, each grain is represented by a unique non-conserved order parameter, $\eta_i$, that equals one inside the corresponding grain, and drops smoothly across the diffuse interface to become zero elsewhere in the domain.

The evolution of each grain's order parameter, $\eta_i$, is governed by the Allen-Cahn equation:
\begin{equation}
\frac{\partial \eta_i}{\partial t} = -L \frac{\delta F}{\delta \eta_i},
\end{equation}
where $L$ is the order-parameter mobility, $F$ is the total free energy of the system, and $\delta F/\delta \eta_i$ is the variational derivative of the free energy with respect to $\eta_i$. The total free energy, $F$, is the integral of the sum of the local free energy density, $f_{\text{loc}}$, and the gradient energy density over the whole domain:
\begin{equation}
F = \int_V \left( f_{\text{loc}} + \frac{\kappa}{2} \sum_{i=1}^N |\nabla \eta_i|^2 \right) \, dV,
\end{equation}
where $\kappa$ is the gradient energy coefficient, $N$ is the total number of grains, and $V$ denotes the volume of the system. The local free energy density, $f_{\text{loc}}$, is given by \cite{Moelans2008}:
\begin{equation}
f_{\text{loc}} = \mu \left[ \sum_{i=1}^N \left( \frac{\eta_i^4}{4} - \frac{\eta_i^2}{2} \right) + \frac{3}{2} \sum_{i=1}^N \sum_{j>i}^N \eta_i^2 \eta_j^2 + \frac{1}{4} \right],
\end{equation}
where $\mu$ is the free energy weight. The model parameters $L$, $\kappa$, and $\mu$ are related to the GB energy, $\gamma$, the diffuse GB width, $w$, and the GB mobility, $M$, through the following expressions \cite{Moelans2008}:
\begin{equation}
L = \frac{4}{3} \frac{M}{w}, \quad \mu = 6 \frac{\gamma}{w}, \quad \kappa = \frac{3}{4} \gamma w.
\label{Eq:Relations}
\end{equation}
All these expressions assume thermal equilibrium, constant molar volume, isotropic GB properties, and symmetric phase-field profiles \cite{Moelans2008}. A symmetric profile satisfies the relation $\eta_i(x) = 1 - \eta_i(-x)$ under the assumption that the central plane of the diffuse GB interface is located at $x = 0$, where $x$ is the coordinate perpendicular to the GB \cite{Moelans2008}. Relations in \cref{Eq:Relations} allow the adjustment of model parameters, i.e., $L$, $\kappa$, and $\mu$, so that the given GB energy and GB mobility are reproduced independently of the diffuse GB width, while letting the diffuse GB width be determined solely based on numerical and computational considerations. That is, in Moelans \textit{et al.}'s grain growth model, the GB width is a numerical rather than a physical parameter.

It is worth discussing the relationship between the physical and numerical pictures of grain growth. GBs in polycrystalline materials are almost planar surfaces (or straight lines in 2D), with abrupt discontinuities at triple junctions, where high local curvature is concentrated \cite{Gottstein2000, Ko2007}. That is, in the diffuse-interface grain growth model \cite{PhaseFieldReview}, triple junctions can be regarded as high-curvature GBs, with no drag force associated with them. The solution procedure in phase-field models is based on free energy minimization \cite{PhaseFieldReview}. By solving the Allen-Cahn equations, the system progressively reduces its free energy. Triple junctions exhibit steeper order parameter gradients, leading to higher energy penalties that drive the system to eliminate these high energy regions (especially if the dihedral angles are far from the equilibrium value of 120$^\circ$ \cite{Abbaschian2025}), ultimately promoting grain growth.

In the MOOSE framework, the Allen-Cahn equations are solved using the finite element method (FEM). In phase-field modeling of polycrystalline grain growth, assigning a unique order parameter to each grain enables accurate tracking but significantly increases computational expense. Conversely, using a single order parameter to represent multiple grains reduces computational cost but introduces artifacts, such as nonphysical grain coalescence when grains associated with the same order parameter meet. To overcome this challenge, the Grain Tracker algorithm~\cite{Permann2016} is employed. This approach permits multiple grains, referred to as \textit{features}, to be associated with a single order parameter, while dynamically remapping the grain-to-order-parameter assignments to prevent nonphysical merging.

% Grain growth simulations were conducted in both two- (2D) and three-dimensional (3D) domains. The 2D simulation domain spans 2500~nm along the $x$- and $y$-axes, while the 3D domain extends 1000~nm in each spatial direction. Initial microstructures were generated using Voronoi tessellation, yielding 500 grains in 2D and 100 grains in 3D. These correspond to average initial grain sizes of $D_0 = 126.16$~nm (2D) and $D_0 = 267.30$~nm (3D). It should be mentioned that the 2D samples nearly replicate the nm-sized thin films prepared by Protasova and Sursaeva \cite{Protasova2001} to study normal grain growth in Al. Simulations were performed over the temperature range 1500--2000~K, in 100~K increments. The initial mesh resolution was set to $120 \times 120$ in 2D and $15 \times 15 \times 15$ in 3D, with uniform refinement applied in both cases. The diffuse GB width, $w$, was fixed at 14~nm for all simulations. Periodic boundary conditions were enforced in all spatial directions. Order parameters were implemented as constant monomial variables. Preliminary tests showed that 20 order parameters in 2D and 25 in 3D are sufficient to resolve grain morphology while maintaining computational efficiency within the Grain Tracker framework. Adaptive time stepping and mesh refinement were employed to ensure accuracy and stability. The simulations began with an initial time step of $10^{-4}$~s and proceeded for 1000 time steps.

Grain growth simulations were conducted in 2D domains that span 2500~nm along the $x$- and $y$-axes. Initial microstructures were generated using Voronoi tessellation, yielding 500 grains in 2D, which correspond to an average initial grain size of $D_0 = 126.16$~nm. It should be mentioned that the 2D samples nearly replicate the nm-sized thin films prepared by Protasova and Sursaeva \cite{Protasova2001} to study normal grain growth in Al.

Simulations were performed over the temperature range 1500--2000~K, in 100~K increments. The initial mesh resolution was set to $120 \times 120$. Uniform refinement of level 2 is applied to the initial mesh. The diffuse GB width, $w$, was fixed at 14~nm for all simulations. Periodic boundary conditions were enforced in all spatial directions. Order parameters were implemented as constant monomial variables. Preliminary tests showed that 20 order parameters in 2D are sufficient to resolve grain morphology while maintaining computational efficiency within the Grain Tracker framework. Adaptive time stepping and mesh refinement were employed to ensure accuracy and stability. The simulations begin with an initial time step of 1~s and proceed for $10^4$ s.

At each time step, the number of grains is recorded, and the average grain size is computed. Grain sizes (or areas in 2D) are extracted using a Vector Postprocessor called the Feature Volume Vector Postprocessor which reads feature data from the Grain Tracker and outputs the area $A_i$ of each grain $i$ at every time step (Note that in MOOSE nomenclature, the term feature here denotes a grain). The average grain area is calculated as:
\begin{equation}
\bar{A}(t) = \frac{1}{N(t)} \sum_{i=1}^{N(t)} A_i,
\label{Eq:Aav}
\end{equation}
where $N(t)$ is the number of grains at time $t$. The average grain diameter $D(t)$ is then obtained from the relation:
\begin{equation}
D(t) = 2 \left( \frac{\bar{A}(t)}{\pi} \right)^{1/2}.
\label{Eq:Dav2D}
\end{equation}
Grain sizes/areas at each time step are also used to assess the evolution of the grain size distribution and to verify the occurrence of normal grain growth. In such growth, the normalized grain size distribution becomes time-invariant and closely follows the Hillert distribution~\cite{Hillert1965}:
\begin{equation}
f(u) = (2e)^d \cdot \frac{d \cdot u}{(2-u)^{2+d}} \exp\left(-\frac{2d}{2-u}\right), \quad 0 \leq u \leq 2,
\end{equation}
where $u = D / \bar{D}$ is the grain size normalized by the average grain size, and $d$ = 2 or 3 indicates the spatial dimension. The Hillert distribution has no free parameters and cannot be directly fitted to data~\cite{Schule1996}. To analyze simulation results, we instead employed a Hillert-like distribution with tunable parameters:
\begin{equation}
f(u) = C u^m \exp(- b u^n), \quad 0 \leq u \leq 2.25,
\label{Eq:Hillert}
\end{equation}
where $C$, $b$, $m$, and $n$ are fitting constants. This function retains the essential qualitative features of the Hillert distribution, e.g., its positive skewness, while offering flexibility for empirical fitting. The upper limit of our distribution has been increased to $u$ = 2.25 based on the observation that a distribution consistent with experimental measurements usually has 99.9\% of the grains below $u$ = 2.2 \cite{Breithaupt2021}.

In the following section, we present an uncertainty analysis to evaluate the impact of uncertainties in the GB mobility prefactor, GB mobility activation energy, and GB energy on the final grain structure. In contrast, performing a similar analysis for uncertainties in the initial grain size distribution is more complex, as this distribution represents a topological characteristic of the simulation rather than a simple numerical input. Specifically, the probability distribution of the initial grain sizes cannot be readily quantified. To approximate the influence of this uncertainty, we analyze grain growth across ten different initial microstructures generated using ten distinct random seeds. The reported grain growth is then taken as the average over these simulations.

\subsection{Uncertainty quantification and sensitivity analysis}

Understanding how uncertainties in input parameters influence grain growth predictions in phase-field models is essential for assessing model robustness and identifying dominant factors. To this end, we implement a surrogate-assisted framework that combines uncertainty propagation, dimensionality reduction, Gaussian process regression, and global sensitivity analysis. The uncertainty quantification and sensitivity analysis calculations are conducted using in-house Python scripts. The grain growth observations are treated as noise-free. Hence, uncertainty is attributed exclusively to model parameters. We focus on three key uncertain inputs: the mobility prefactor $M_0$, the activation energy $Q_M$, and the GB energy $\gamma$.

The mobility $M$ is estimated from grain growth experimental data \cite{Ronchi1975} via regression (\cref{sec:GBMob}), yielding maximum likelihood estimates and a covariance matrix for $M_0$ and $Q_M$. The prefactor $M_0 = e^b$ is modeled as log-normally distributed, where $b \sim \mathcal{N}(\mu = -33.82, \sigma^2 = 0.82^2)$. The activation energy $Q_M$ follows a normal distribution $\mathcal{N}(\mu = 0.89 \text{ eV}, \sigma^2 = 0.11^2 \text{ eV}^2)$ (see \cref{sec:GBMob}). The GB energy, $\gamma$, is modeled as a normal distribution. Its mean is obtained from a correlation fit to MD data of 27 GB energies (\cref{Eq:gamma}), while the standard deviation is set to 29.6\% of the mean, based on a statistical analysis of the same dataset. All uncertain inputs are sampled from their prescribed distributions using inverse-CDF (PPF) transforms with an enforced $\pm 3 \sigma$ cutoff in standard-normal space. In addition, the activation energy $Q_M$ and grain-boundary energy $\gamma$ are modeled as truncated normal distributions constrained to positive values only.

Our quantity of interest (QoI) is the average grain size $D(t)$. Because $D(t)$ evolves over time, the QoI is inherently multivariate and can be represented as a vector-valued function. Each element in this vector corresponds to a time-resolved scalar output, requiring specialized methods for uncertainty quantification and sensitivity analysis that account for temporal dynamics. For this purpose, we utilize a methodology outlined in detail in \cite{Yaseen2023, Yaseen2025, Nasr2025global, Nasr2025mcre}. Here, we only give a brief overview.

\begin{figure}[h!]
  \centering
  \includegraphics[width=0.7\textwidth]{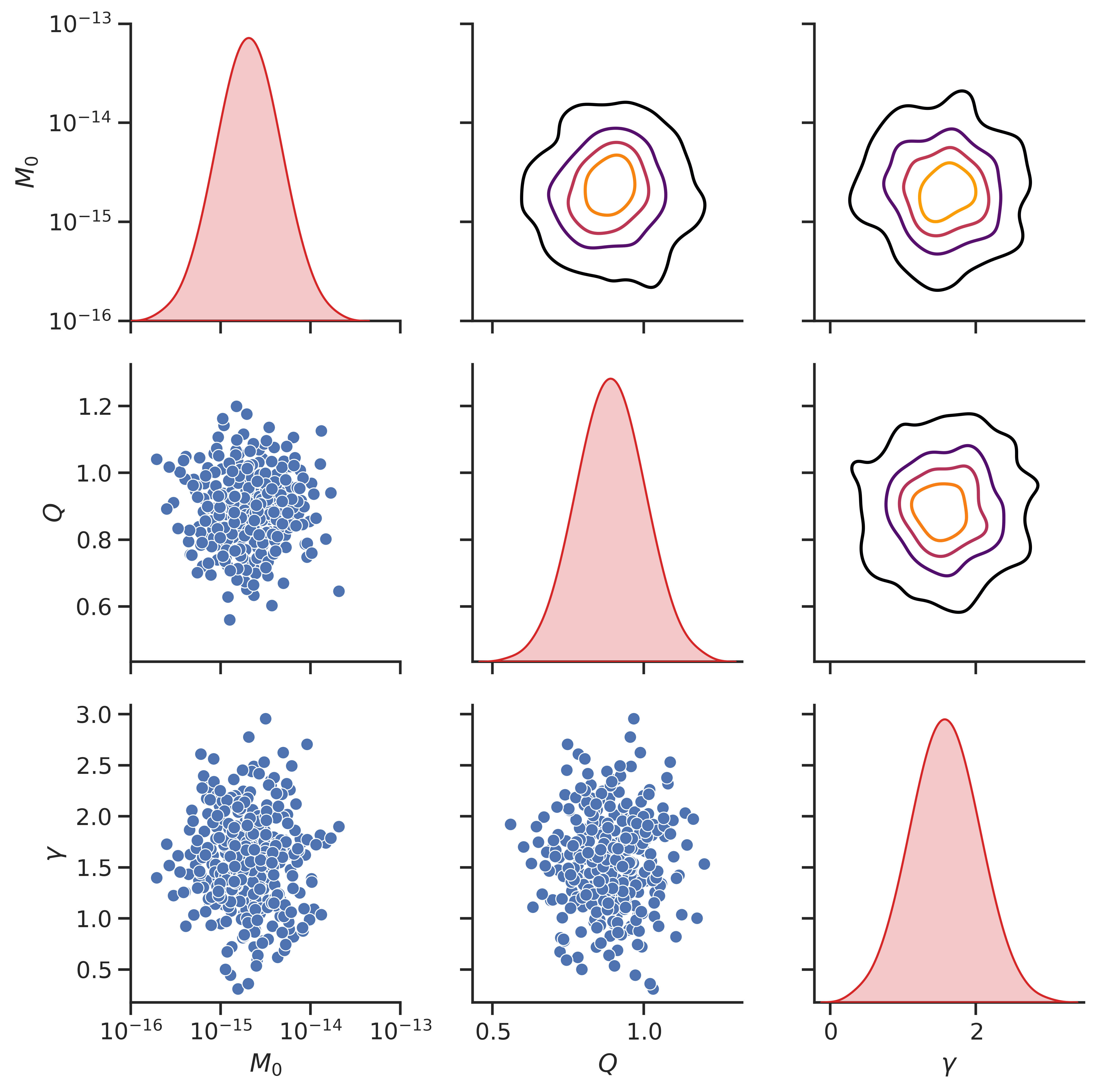}
  \caption{Input parameter samples and pairwise correlations used in the UQ analysis. Samples are drawn for three uncertain quantities: the natural logarithm of the GB mobility prefactor $\ln(M_0)$ (where $M_0$ is in m$^4$/J$\cdot$s), the mobility activation energy $Q_M$ (eV), and the GB energy $\gamma$ (J/m$^2$). Diagonal panels show their marginal distributions, while off-diagonal panels show correlations among parameters.} 
  \label{Fig:SampleDist}
\end{figure}

To propagate uncertainty through the phase-field model, we generate 330 parameter samples using Latin hypercube sampling (LHS)~\cite{McClarren2018} with a maximin criterion and 500 optimization iterations, ensuring good coverage of the input space (\cref{Fig:SampleDist}). For each sample, we perform the MOOSE-based phase-field simulations as described in \cref{Sec:PhaseField}. Due to the high computational cost of these simulations, direct sensitivity analysis on the full time-dependent model outputs is infeasible. Accordingly, the grain size time series are first preprocessed prior to surrogate construction: The data are log-transformed as $\ln(D)$ so that early-time variations do not appear negligible compared to late-time variations. All simulation outputs are mapped onto a common temporal grid to accommodate variations in adaptive time-stepping. Missing values at specified grid points are forward-filled using the last available observation. To reduce temporal redundancy and computational cost, the time series are downsampled by retaining every 20th time step after skipping the initial 5 transient steps. The analysis is truncated at 5000 s to maintain sufficient grain count for reliable statistics.

To reduce the dimensionality of the resulting time-dependent QoI, we then apply principal component analysis (PCA). Prior to PCA, the log-transformed grain size data are standardized independently at each time step to zero mean and unit variance across runs. PCA transforms the correlated time series into a set of uncorrelated principal components (PCs) that preserve 98\% of the total variance, which are then modeled with Gaussian process surrogates.

$M_0$ is log-transformed to $\ln(M_0)$ to account for its log-normal distribution and reduce skewness, while $Q_M$ and $\gamma$ remain in their original scale. All three inputs are then standardized to zero mean and unit variance. Independent single-output Gaussian process regression models are trained for each of the PCs using the \verb|GPflow| library~\cite{GPflow2017, GPflow2020multioutput}. A Mat\'ern-5/2 kernel is employed to provide twice-differentiable predictive mean functions.

The GP model decomposes the observed PC scores into a smooth underlying function and observational noise:
\begin{equation}
y = f(\mathbf{x}) + \epsilon, \quad \epsilon \sim \mathcal{N}(0, \sigma_n^2),
\end{equation}
where $f(\mathbf{x})$ represents the deterministic relationship between input parameters and PC scores, and $\sigma_n^2$ is the noise variance. Since the phase-field simulations are deterministic, $\sigma_n^2$ is expected to be small. It is therefore regularized by a $\mathrm{Gamma}(2, 200)$ prior, which places a weakly informative constraint with mean $\mathbb{E}[\sigma_n^2] = 2/200 = 0.01$ in standardized PC space, preventing the optimizer from inflating $\sigma_n^2$ to absorb signal. Model hyperparameters are optimized using the Adam optimizer with a learning rate of 0.001, employing early stopping with a patience of 200 iterations and a minimum improvement threshold of $10^{-4}$ to prevent overfitting.

The GP surrogate accuracy is assessed using in-sample $R^2$ and predictive $Q^2$, computed under repeated five-fold cross-validation with five repeats (25 train-test splits total). The trained surrogates predict PC scores for new parameter samples, which are then inverted via PCA to reconstruct full time-dependent grain size profiles. The Python library \verb|scikit‑learn| \cite{scikit-learn} is used in this study for dimensionality reduction via PCA, standardization of input parameters and PC scores, and rigorous model validation through repeated cross-validation.

To quantify the influence of each uncertain input over time, we perform a variance-based global sensitivity analysis using Sobol indices~\cite{Sobol1993, Sobol2001, Saltelli2010}. First-order ($S_1$) and total-effect ($S_t$) indices are computed to assess both individual and interactive contributions of parameters to the variance of the QoI. These indices are calculated at each time step, enabling time-resolved sensitivity analysis. The Saltelli sampling scheme, implemented via the \verb|SALib| package~\cite{SALib1, SALib2}, is employed to estimate first-order and total-order Sobol indices with a base sample size of $N = 2^{13} = 8{,}192$, requiring $N(d+2) = 40{,}960$ model evaluations in the three-dimensional ($d$ = 3) input space. The GP surrogate replaces the computationally intensive phase-field model, allowing efficient and accurate estimation of Sobol indices across the time domain. Point estimates of the indices are obtained from the GP predictive mean, with confidence intervals computed using the bootstrap resampling method implemented in \verb|SALib|.

\begin{figure}[h!]
    \centering
    \begin{tikzpicture}[
        node distance=1cm and 1.8cm,
        every node/.style={draw, align=center, font=\small, fill=blue!10, text=black, minimum height=1cm},
        arrow/.style={-Stealth, thick, color=blue!70},
        precursorsArrow/.style={-Stealth, thick, color=red!70},
        powerDensityArrow/.style={-Stealth, thick, color=green!70}
    ]

    % Nodes
    \node (perturb) [fill=red!20] {Perturb Input Parameters \\ ($\ln(M_0)$, $Q_M$, $\gamma$)};
    \node (moose) [below=0.8cm of perturb, fill=green!30, minimum width=2.5cm] {MOOSE \\ (Phase Field Module)};
    \node (qoi) [below=0.8cm of moose, fill=cyan!20, minimum width=3cm] {Calculate Quantity \\ of Interest (QoI)};
    \node (meanvar) [below=0.8cm of qoi, fill=gray!20, minimum width=3cm] {Compute Mean/Variance \\ of QoI with Time};
    \node (pca) [left=1.8cm of qoi, fill=yellow!20, minimum width=2.8cm] {Perform PCA \\ for QoI};
    \node (gp) [below=0.8cm of pca, fill=orange!20, minimum width=2.8cm] {Train the Gaussian \\ Process (GP) on PCs};
    \node (gen_gp) [below=0.8cm of gp, fill=purple!20, minimum width=2.8cm] {Generate Samples \\ Using the GP};
    \node (inv_pca) [right=1.8cm of gen_gp, xshift=-0.5cm, fill=red!20, minimum width=2.8cm] {Inverse PCA \\ to Retrieve Time Series};
    \node (sobol) [below=0.8cm of inv_pca, fill=orange!20, minimum width=2.8cm] {Compute Sobol Indices \\ at Each Time Step};

    % Arrows
    \draw[arrow] (perturb) -- (moose);
    \draw[arrow] (moose) -- (qoi);
    \draw[arrow] (qoi) -- (meanvar);
    \draw[arrow] (qoi.west) -- (pca.east);
    \draw[arrow] (pca.south) -- (gp.north);
    \draw[arrow] (gp.south) -- (gen_gp.north);
    \draw[arrow] (gen_gp.east) -- (inv_pca.west);
    \draw[arrow] (inv_pca.south) -- (sobol.north);

    \end{tikzpicture}
    \caption{Computational methodology for time-dependent Sobol indices.}
    \label{Fig:GSA}
\end{figure}
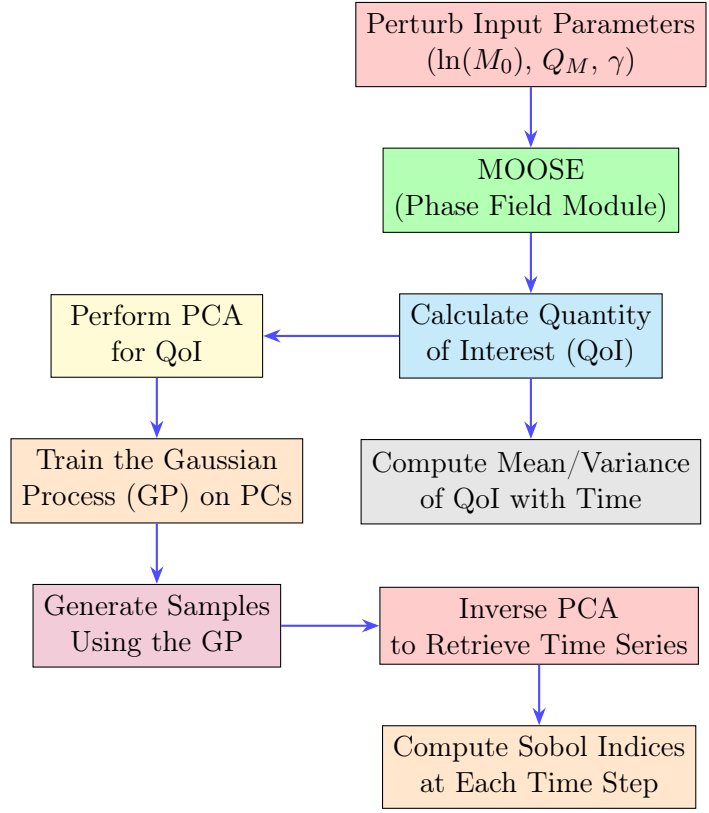

This surrogate-assisted methodology, summarized in \cref{Fig:GSA}, enables efficient uncertainty quantification and sensitivity analysis for multivariate, time-resolved outputs. 

\section{Results}

\subsection{Grain-boundary energy}
\label{Sec:GBE}

\begin{figure}[h!]
\centering
\begin{subfigure}{0.3\textwidth}
    \includegraphics[width=\textwidth]{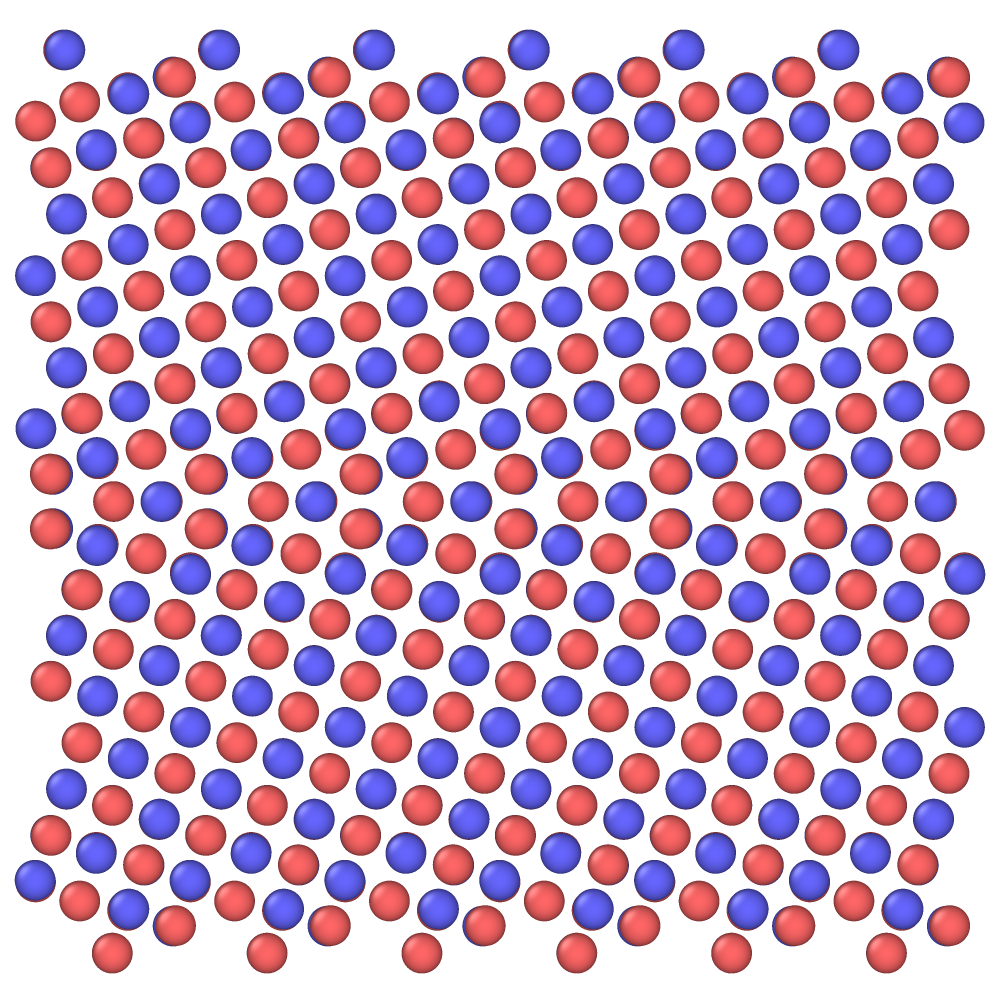}
    \caption{Tseplyaev potential}
    \label{Fig:ADP0K}
\end{subfigure}
\hspace{1em}
\begin{subfigure}{0.3\textwidth}
    \includegraphics[width=\textwidth]{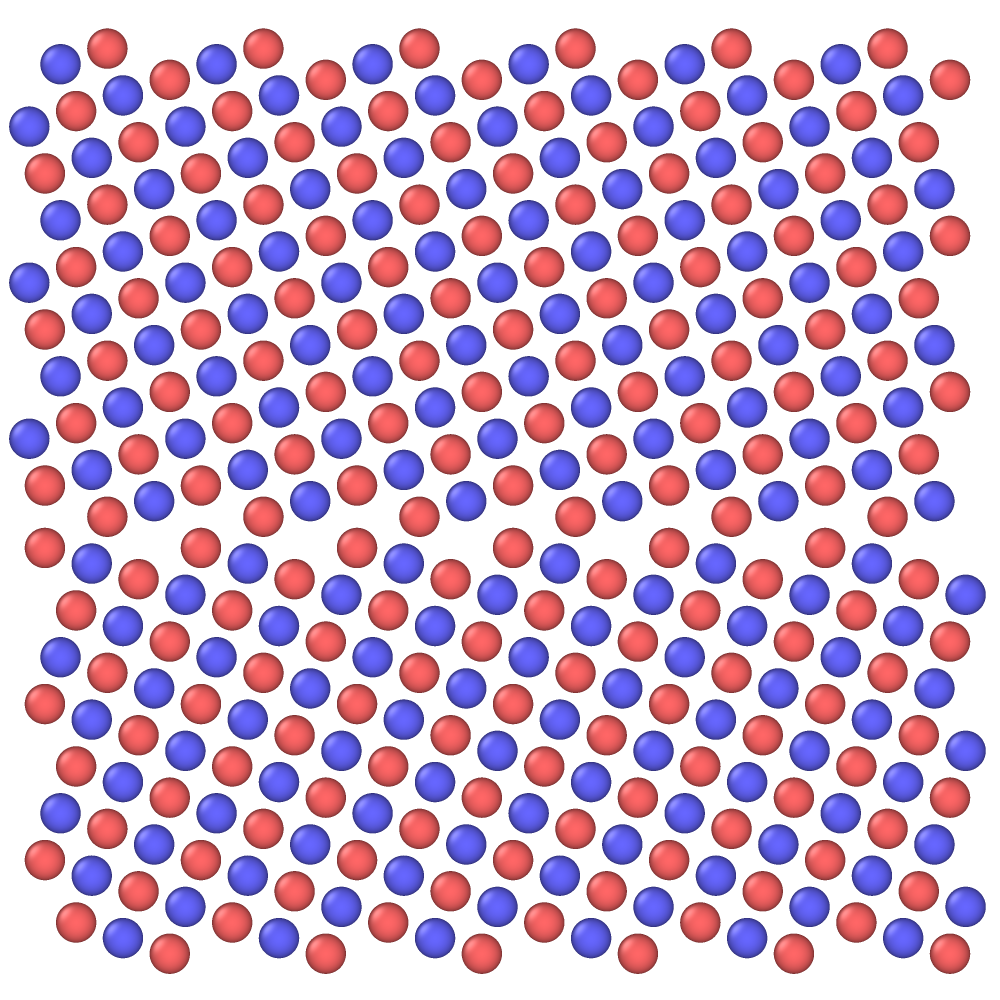}
    \caption{Kocevski potential}
    \label{Fig:EAM0K}
\end{subfigure}
\caption{(Color online) Top view along the $z$-axis of the same configuration of the (310) GB plane after 0~K minimization using \textbf{(a)} the Tseplyaev potential, and \textbf{(b)} the Kocevski potential. U atoms are colored in red, and N atoms are colored in blue.}
\label{Fig:GBStructure}
\end{figure}

Representative minimized bicrystal GBs from the Tseplyaev and Kocevski potentials are shown in \cref{Fig:GBStructure}. After 0~K minimization of the bicrystals using the Kocevski potential, a void space is still observed between the two grains. That is, the GB doesn't ``close'' after minimization (\cref{Fig:EAM0K}). This is due to the inability of the Kocevski potential to simulate metallic U, which leads to mutual repulsion among U atoms. For the Tseplyaev potential, the GB closes after 0~K minimization as expected (\cref{Fig:ADP0K}), because it was demonstrated that the Tseplyaev potential can simulate metallic U \cite{AbdulHameed2024} and expresses no anomalous effects related to U-U repulsion. However, the Kocevski potential is still utilized to calculate GB energies at 0~K for completeness.

\begin{figure}[h!]
\centering
\begin{subfigure}{0.48\textwidth}
    \includegraphics[width=\textwidth]{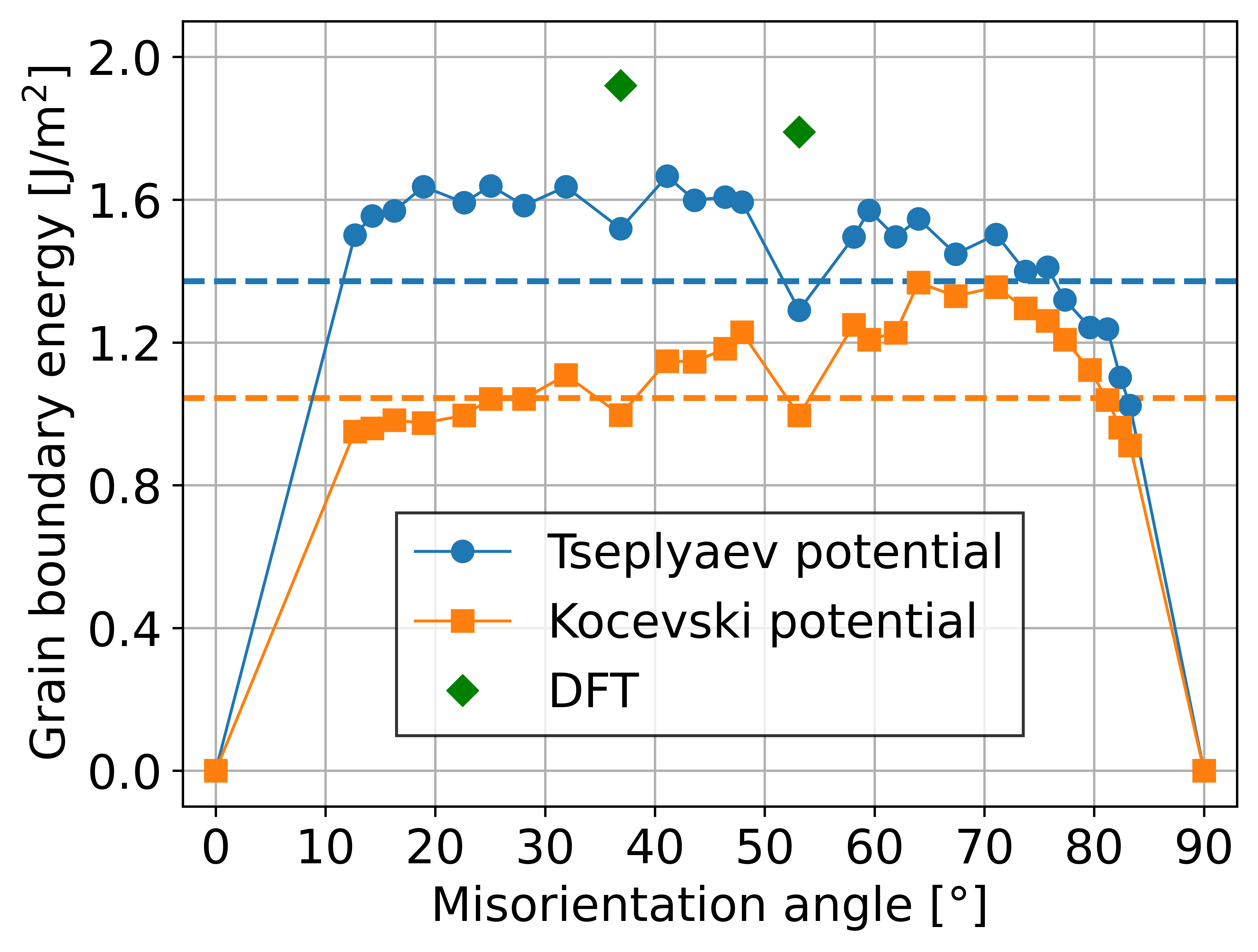}
    \caption{}
    \label{Fig:GB-0}
\end{subfigure}
\hfill
\begin{subfigure}{0.48\textwidth}
    \includegraphics[width=\textwidth]{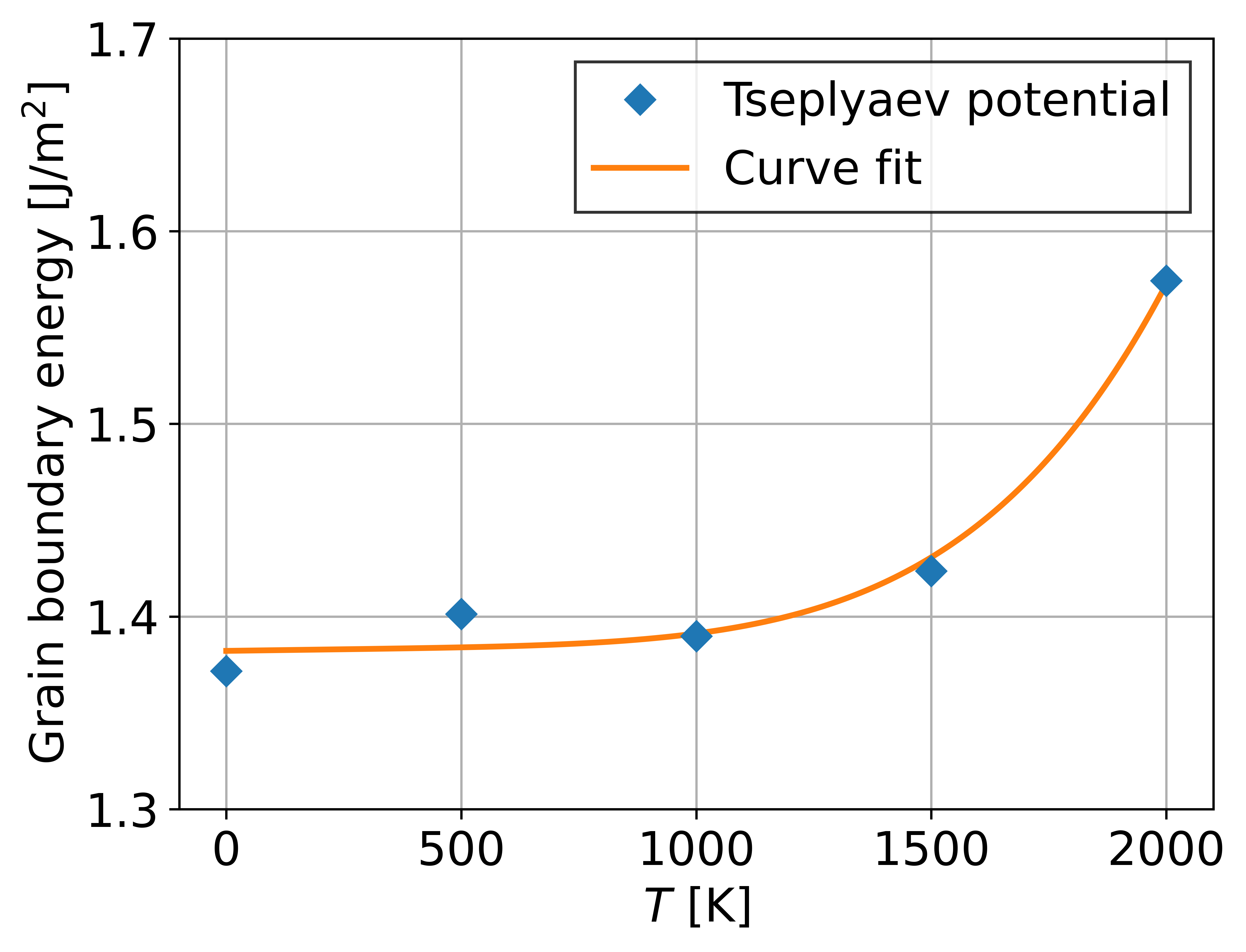}
    \caption{}
    \label{Fig:GB-1}
\end{subfigure}
\caption{\textbf{(a)} Variation of the GB energy with misorientation angle at 0~K, as predicted by both potentials. The DFT values are from \cite{Wang2023}. The dashed lines correspond to an average of the GB energy overall misorientation angles. \textbf{(b)} Variation of the average GB energy with temperature as predicted by the Tseplyaev potential.}
\label{Fig:GB}
\end{figure}

The GB energy as a function of misorientation angle at 0~K and the GB energy as a function of temperature are shown in \cref{Fig:GB}. From \cref{Fig:GB-0}, the Kocevski potential consistently predicts smaller GB energies than those predicted by the Tseplyaev potential at 0~K. This stems from the fact that U-U repulsion shifts pair distances to the tail of the potential energy versus distance curve, where the interaction energy is usually small. Because a large number of the GB planes with misorientation angles within the range of 10$^\circ$--30$^\circ$ form coincidence site lattices (CSLs), with $\Sigma$ values ranging from 13 to 41 (\cref{Tab:GBPlanes}), they display very close GB energies nearly independent of the misorientation angle. The situation is different for GB planes with misorientation angles in the range of 70$^\circ$--90$^\circ$, which do not form CSLs and show the trend expected for low-angle GBs \cite{Abbaschian2025}.

The (310) and (210) GB planes, with misorientation angles of 36.87$^\circ$ and 53.13$^\circ$, respectively, show remarkably low energies. These planes have the smallest $\Sigma$ value possible for a tilt GB, i.e., $\Sigma$ = 5. For GBs with low $\Sigma$ numbers, a large fraction of the atoms in the GB plane are positioned exactly as they would be in a perfect lattice, which leads to a smaller GB energy \cite{Cai2016}. Note that in \cref{Fig:GB-0}, the GB energy predicted by the Tseplyaev potential is close to the DFT values, whereas that predicted by the Kocevski potential is largely underestimated. It should be mentioned that we calculated the energy of a few symmetric tilt GBs at 0~K using much larger supercells and found the GB energy to be independent of the supercell size.

Because the Kocevski potential is not able to properly simulate GBs, the finite-temperature calculation of the GB energies was performed using only the Tseplyaev potential. The variation of the GB energy with misorientation angle predicted at finite temperatures (not shown) is nearly identical to that at 0~K (\cref{Fig:GB-0}). An average GB energy over all of the misorientation angles at each temperature is calculated and plotted in \cref{Fig:GB-1}. The Tseplyaev potential also predicts a nearly temperature-independent GB energy up to about 1000~K and then an increase in GB energy at higher temperatures. The data points in \cref{Fig:GB-1} have been fitted to a polynomial function:
\begin{equation}
\gamma = 1.382 + 3.279 \times 10^{-6} \ T + 5.748 \times 10^{-18} \ T^5 ~ (\text{J}/\text{m}^2)
\label{Eq:gamma}
\end{equation}
with $R^2$ = 0.983. Because of the large power of $T$ in \cref{Eq:gamma}, the fitted correlation should be used with caution outside the fitted range, i.e., at $T > 2000$ K. The standard deviation of the averaged 27 GB energies is found to be about 30\% at all temperatures. A GB energy that increases with temperature has been observed for U$_3$Si$_2$ by Beeler \textit{et al.} \cite{Beeler2019}, and is often correlated with increased disorder in the GB region \cite{Frolov2012}. This disorder likely stems from the enhanced bulk diffusivity of the nitrogen sublattice, which remained zero until near 2000 K and then began to increase, as observed in our previous work \cite{AbdulHameed2024c}.

\subsection{Grain-boundary mobility}
\label{sec:GBMob}

\begin{figure}[h!]
  \centering
  \includegraphics[width=0.6\textwidth]{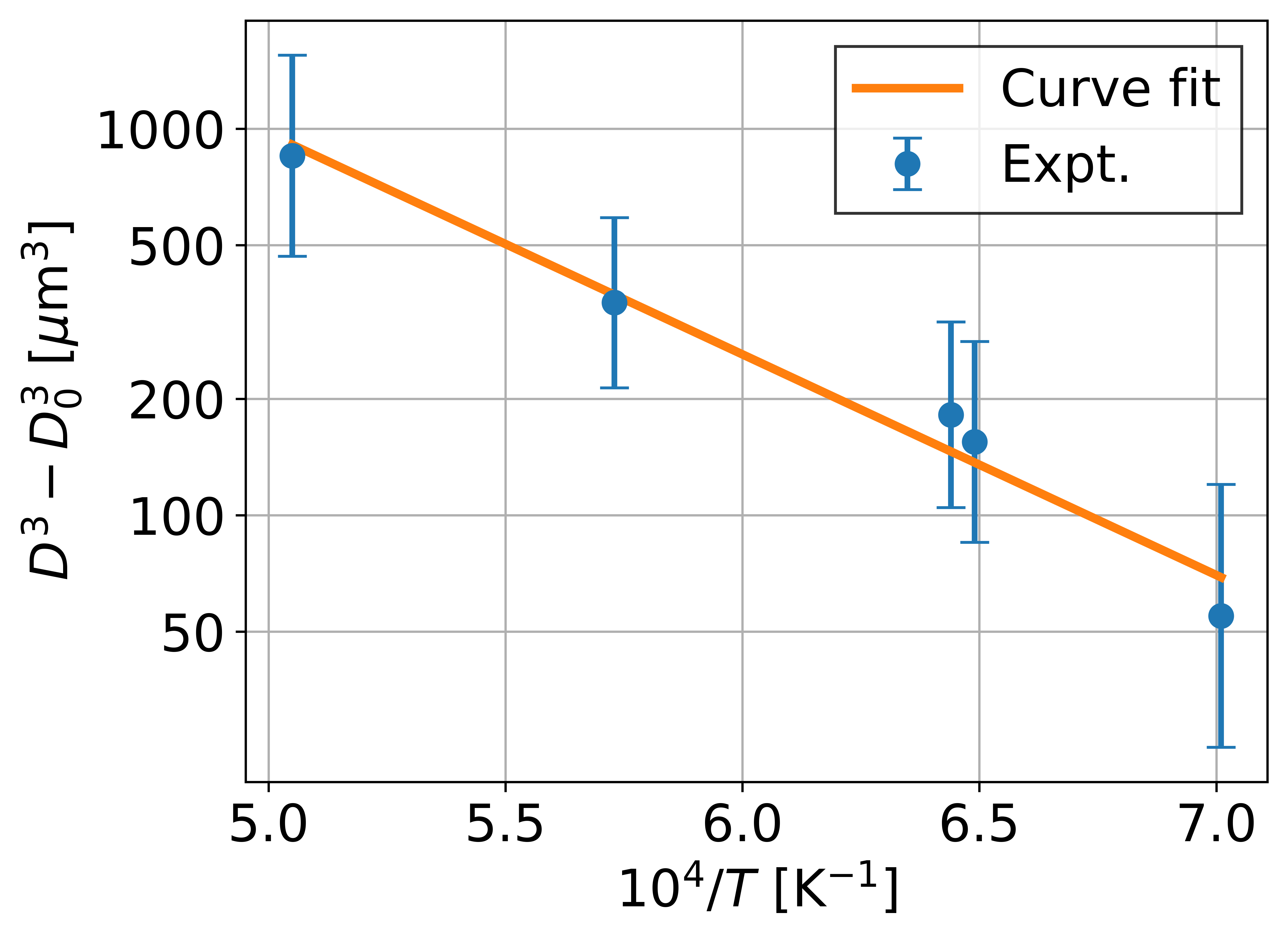}
  \caption{Grain growth Arrhenius plot for (U,Pu)N pellets heated in a thermal gradient of 1000~K/cm for an annealing time of 500 hours. The data is taken from Ronchi and Sari \cite{Ronchi1975}. The data were fitted to the equation $D^3 - D_0^3 = K t$, with $Q_K = 1.13 \pm 0.11$~eV, and $K_0 = 3.856 \times 10^{-19}$~m$^3$/s.} 
  \label{Fig:GGUN}
\end{figure}

GB mobility, $M$, is a fundamental parameter governing grain growth kinetics in nuclear fuel materials. Its determination requires careful experimental measurement, theoretical analysis, and/or modeling. In the literature, only one experimental study has investigated grain growth in actinide nitride fuels. Ronchi and Sari \cite{Ronchi1975} conducted measurements on (U,Pu)N pellets subjected to annealing at 1400--2000~K for 500 hours under thermal gradients of approximately 1000~K/cm. The pellets had initial grain size $D_0 = 6~\mu$m and porosity of approximately 15~vol.\%, with 55\% of this porosity located at grain boundaries. Their measurements followed the empirical growth law:
\begin{equation}
D^3 - D_0^3 = K(T)\,t,
\label{Eq:GGexp}
\end{equation}
where the rate constant $K(T)$ exhibits Arrhenius temperature dependence, as shown in \cref{Fig:GGUN}. The cubic growth exponent $n=3$ deviates from the quadratic behavior $n=2$ expected for drag-free curvature-driven grain growth, indicating the presence of a resistive mechanism commonly attributed to pore drag in porous ceramics.

We are confronted with two challenges in extracting intrinsic GB mobility from these measurements. First, the grain growth was observed under a temperature gradient rather than isothermal conditions. Second, the grain size exponent is $n=3$ instead of the ideal $n=2$ case expected for drag-free growth. Tonks \textit{et al.} \cite{Tonks2014} and Bai \textit{et al.} \cite{Bai2015} showed that in UO$_2$, temperature gradients have minimal impact on the \textit{overall} grain growth behavior when compared to the curvature driving force. Indeed, temperature gradients do affect temperature-dependent GB mobility, causing GBs to migrate quickly in hotter areas and slowly in cooler areas. As a result, temperature gradients influence \textit{local} grain growth but do not change the average grain growth behavior. Given that UN exhibits higher thermal conductivity than UO$_2$, its operational temperature gradient is reduced to \nicefrac{1}{5}--\nicefrac{1}{2} that of UO$_2$ \cite{Ronchi1975,Olander2017}, further decreasing the significance of the temperature gradient as a driving force for grain growth in UN. Therefore, the measurements from Ronchi and Sari are assumed to be approximately indicative of isothermal grain growth.

For the second challenge, we present a methodology to extract the drag-free GB mobility, $M$, from the cubic grain growth kinetics. The procedure consists of three steps: (\textit{i}) determination of the effective mobility from experimental kinetics, (\textit{ii}) construction of a mechanistic pore-drag model with partial coupling, and (\textit{iii}) global regression to extract intrinsic mobility parameters.

\subsubsection*{Effective mobility from experimental kinetics}

The temperature-dependent growth constant in \cref{Eq:GGexp} follows an Arrhenius relation:
\begin{equation}
K(T) = K_0 \exp\!\left(-\frac{Q_K}{k_B T}\right).
\end{equation}
Taking the logarithm yields a linear form:
\begin{equation}
\ln K = \ln K_0 - \frac{Q_K}{k_B}\frac{1}{T},
\end{equation}
which is fitted by linear regression in $1/T$ to obtain the activation energy $Q_K$ and pre-exponential factor $K_0$, along with their c ovariance matrix. From the experimental data of Ronchi and Sari, we extract $Q_K = 1.13 \pm 0.11$~eV and $K_0 = 3.856 \times 10^{-19}$~m$^3$/s.

The cubic exponent $n=3$ suggests the presence of a resistive force, likely pore drag. Powers and Glaeser~\cite{Powers1998} proposed that such non-parabolic kinetics can be rationalized by assuming the growth constant depends on the instantaneous grain size. Specifically, they introduced an effective mobility $M_{\mathrm{eff}}(T)$ such that:
\begin{equation}
D^3 - D_0^3 = 3 \alpha M_{\mathrm{eff}}(T) \gamma(T) D(T) t,
\label{Eq:PG_formulation}
\end{equation}
where $\alpha \approx 1$ is a geometric constant for 3D grains~\cite{Tonks2021} and $\gamma(T)$ is the GB energy.

Equating \cref{Eq:GGexp,Eq:PG_formulation} and rearranging yields the effective mobility:
\begin{equation}
M_{\mathrm{eff}}(T) = \frac{K(T)}{3\alpha\,\gamma(T)\,D(T)}.
\label{Eq:Meff_extraction}
\end{equation}
Here, $\gamma(T)$ is obtained independently from MD simulations (\cref{Sec:GBE}), and $D(T)$ is reconstructed from the experimental grain growth data at each temperature by solving $D = (D_0^3 + K(T)t)^{1/3}$.

\subsubsection*{Mechanistic pore-drag model}

The effective mobility $M_{\mathrm{eff}}$ differs from the intrinsic (drag-free) mobility, $M$, due to the resistive force exerted by pores on migrating GBs. Following the formulation of Tonks \textit{et al.}~\cite{Tonks2021}, which builds on the classical framework of Powers and Glaeser~\cite{Powers1998}, the mobility reduction is expressed as:
\begin{equation}
M_{\mathrm{eff}}(T) = \frac{M(T)}{1+\frac{N_a(T)M(T)}{M_p(T)}},
\label{Eq:drag_relation}
\end{equation}
where $M(T)$ is the intrinsic GB mobility, $N_a(T)$ is the areal number density of pores attached to GBs and participating in drag, and $M_p(T)$ is the pore mobility in units of m$^2$/(J$\cdot$s), with is different from the unit of GB mobility (i.e., m$^4$/(J$\cdot$s)).

To quantify the degree of mobility reduction, we define the dimensionless \textit{drag parameter}:
\begin{equation}
\lambda(T) = \frac{N_a(T)M(T)}{M_p(T)},
\label{Eq:lambda_def}
\end{equation}
which represents the ratio of the GB transport capacity to the pore transport capacity. When $\lambda \ll 1$, pores are highly mobile relative to GBs and drag effects are negligible. When $\lambda \gg 1$, pores are effectively immobile and strongly drag (or possibly pin) GBs.

The \textit{mobility reduction factor} is then defined as:
\begin{equation}
s(T) \equiv \frac{M_{\mathrm{eff}}(T)}{M(T)} = \frac{1}{1+\lambda(T)}.
\label{Eq:reduction_factor}
\end{equation}

This factor quantifies the fractional reduction in GB mobility due to pore drag, with $s = 1$ indicating no drag (drag-free conditions) and $s \to 0$ indicating strong drag and possibly pinning. $s(T)$ is not assumed \textit{a priori} but is computed mechanistically from the underlying microstructural parameters: $M$, $M_p$, and $N_a$.

\subsubsection*{Areal pore density with partial coupling}

In the intermediate-temperature regime (1400--2000~K), Ronchi and Sari reported three particularly relevant observations on microstructural evolution: (\textit{i}) pore number density reaches an approximate plateau at about $n \approx 1.35 \times 10^{14}$~m$^{-3}$, (\textit{ii}) the characteristic pore diameter scales nearly linearly with grain size such that $D_p / D \approx 0.54$, and (\textit{iii}) micrographic evidence shows that some pores are left behind by advancing GBs rather than remaining continuously attached. These observations suggest that pore-GB coupling is partial rather than complete.

To account for partial coupling, we introduce an effective coupling fraction $\phi \in [0,1]$ that represents the fraction of pores that actively participate in drag:
\begin{equation}
N_a(T) = \phi N_p(T),
\end{equation}
where $N_p(T)$ is the total pore number density per GB area (wether attached to GBs or not). The parameter $\phi$ is a lumped effective quantity accounting for two physical effects: (\textit{i}) not all pores in the bulk reside on GBs (some are intragranular), and (\textit{ii}) of those pores that do contact GBs, only a fraction remain attached and migrate with the boundary at any given time (the rest are left behind).

For equiaxed, nearly spherical grains of size $D$, the GB surface area per unit volume is:
\begin{equation}
S_v = \frac{1}{2} \frac{\pi D^2}{\pi D^3 / 6} = \frac{3}{D},
\end{equation}
where the factor of $\nicefrac{1}{2}$ accounts for each GB being shared by two grains. If $n$ is the volumetric pore number density, then the total areal pore number density is:
\begin{equation}
N_p(T) = \frac{n}{S_v} = \frac{nD(T)}{3}.
\label{Eq:Np_areal}
\end{equation}

\subsubsection*{Pore mobility}

For spherical pores migrating via surface diffusion, the pore mobility is given by \cite{Powers1998}:
\begin{equation}
M_p(T) = \frac{D_s(T)\delta_s\Omega}{k_B T\pi r_p^4(T)},
\label{Eq:Mp_pore}
\end{equation}
where $D_s(T)$ is the surface diffusivity, $\delta_s = 0.5$~nm \cite{Tonks2021} is the effective thickness of the surface diffusion layer, $\Omega = a^3/8$ is the atomic volume for UN with NaCl structure (lattice parameter $a = 4.89$~\AA), and $r_p(T)$ is the pore radius. \cref{Eq:Mp_pore} assumes that pores remain spherical during migration.

Direct measurements or atomistic calculations of surface diffusivity in UN are not available. As an approximation, we adopt an Arrhenius fit to the recommended order-of-magnitude estimate for UO$_2$ from Muntaha \textit{et al.}~\cite{Muntaha2024}, which gives $D_s \sim 10^{-9}$~m$^2$/s at 2000~K and $D_s \sim 10^{-17}$~m$^2$/s at 1000~K. This introduces substantial uncertainty in $M_p(T)$, potentially spanning an order of magnitude, which propagates into the pore drag model. Therefore, the extracted intrinsic mobility parameters should be interpreted as order-of-magnitude estimates. 

% However, the global regression procedure performed later partially compensates for uncertainty in the assumed surface diffusivity $D_s(T)$ through the fitted coupling parameter $\phi$. In particular, multiplicative errors in the magnitude of $D_s(T)$ render $\phi$ and $D_s(T)$ correlated, so $\phi$ effectively absorbs systematic bias in the assumed diffusivity magnitude. This compensation is partial: errors in the activation energy of $D_s(T)$ alter the temperature dependence of $M_p(T)$ and cannot be captured by $\phi$. 

% with $\phi$ representing an effective coupling strength that lumps together partial pore-GB attachment and uncertainties in pore transport properties.

Based on the observed scaling relation $D_p/D \approx 0.54$ \cite{Ronchi1975}, we model the pore radius as:
\begin{equation}
r_p(T) = \frac{D_p}{2} \approx 0.27 D(T).
\label{Eq:rp_scaling}
\end{equation}

\subsubsection*{Global regression for intrinsic mobility}

The intrinsic GB mobility follows an Arrhenius relation:
\begin{equation}
M(T) = M_0 \exp\!\left(-\frac{Q_M}{k_B T}\right).
\label{Eq:M_arrhenius}
\end{equation}
Substituting \cref{Eq:Np_areal,Eq:Mp_pore,Eq:rp_scaling,Eq:M_arrhenius} into the drag relation \cref{Eq:drag_relation} gives the forward model for the effective mobility:
\begin{equation}
\hat{M}_{\mathrm{eff}}(T; M_0, Q_M, \phi) = \frac{M_0 \exp \left(- Q_M / (k_B T) \right)}{1 + \frac{\phi \, n \, D(T) \, M_0 \exp \left(- Q_M / (k_B T) \right)}{3M_p(T)}},
\label{Eq:Meff_forward}
\end{equation}
where all temperature-dependent quantities ($D(T)$, $M_p(T)$ via $D_s(T)$ and $r_p(T)$) are evaluated at the experimental temperatures.

The three unknown parameters $(M_0, Q_M, \phi)$ are obtained by minimizing the logarithmic least-squares objective:
\begin{equation}
\min_{M_0, Q_M, \phi} \sum_i \left[\ln \hat{M}_{\mathrm{eff}}(T_i; M_0, Q_M, \phi) - \ln M_{\mathrm{eff}}(T_i)\right]^2,
\label{Eq:objective_lsq}
\end{equation}
subject to the physical constraints:
\begin{equation}
M_0 > 0, \qquad Q_M > 0, \qquad 0 \leq \phi \leq 1.
\end{equation}
The logarithmic objective function ensures approximately uniform weighting in relative error across the temperature range, avoiding overemphasis of high-temperature points where mobilities are largest.

The optimization is performed using the trust-region reflective algorithm \cite{Li2024} with initial guess obtained from the effective mobility Arrhenius fit ($\ln M_{0,\text{initial}} = \ln M_{0,\mathrm{eff}}$, $Q_{M,\text{initial}} = Q_{\mathrm{eff}}$) and a moderate initial coupling fraction ($\phi_\text{initial} = 0.3$). After fitting, the dimensionless drag parameter, $\lambda(T)$, and mobility reduction factor, $s(T)$, at each experimental temperature are computed from the fitted parameters $(M_0, Q_M, \phi)$.

% \begin{equation}
% \lambda(T) = \frac{N_a(T)\,M(T)}{M_p(T)} = \frac{\phi\,n\,D(T)}{3} \frac{M(T)}{M_p(T)},
% \label{Eq:lambda_computed}
% \end{equation}
% \begin{equation}
% s(T) = \frac{1}{1+\lambda(T)}.
% \label{Eq:s_final}
% \end{equation}

The final relation linking the experimentally measured growth constant, $K(T)$, to the intrinsic mobility, $M(T)$, is therefore:
\begin{equation}
K(T) = 3\alpha\,\gamma(T)\,D(T)\,s(T)\,M(T) = 3\alpha\,\gamma(T)\,D(T)\,\frac{M(T)}{1+\lambda(T)},
\label{Eq:K_final_relation}
\end{equation}
which explicitly shows how the phenomenological growth constant depends on both the intrinsic mobility and the mechanistically-computed drag effects.

\subsubsection*{Fitted parameters and diagnostics}

\begin{figure}[h!]
  \centering
  \includegraphics[width=0.6\textwidth]{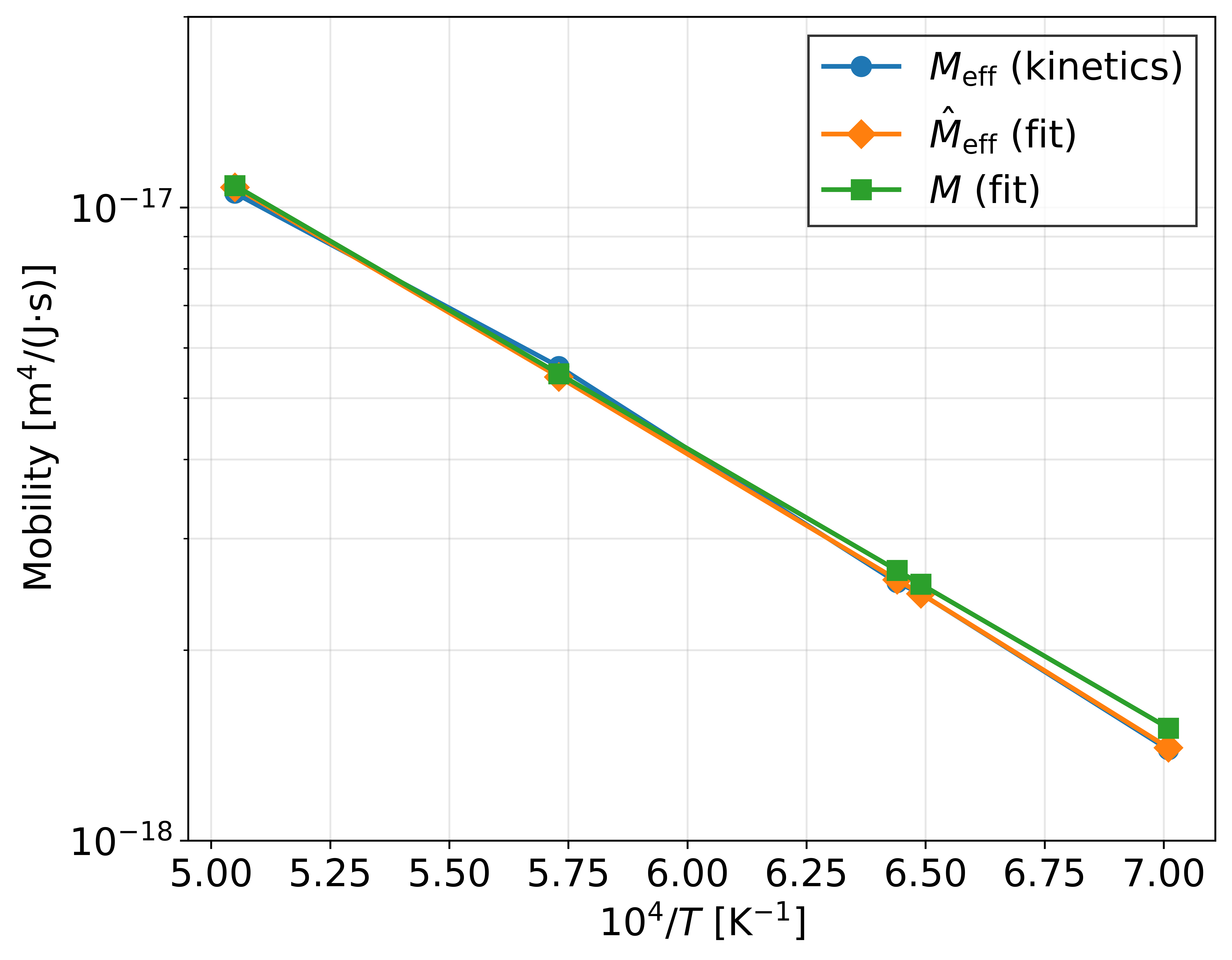}
  \caption{Temperature dependence of grain boundary mobility in uranium mononitride: drag-free intrinsic mobility $M$ (squares), kinetic effective mobility $M_{\mathrm{eff}}$ from experimental data (circles), and model-predicted effective mobility $\hat{M}_{\mathrm{eff}}$ (diamonds). Agreement between $M_{\mathrm{eff}}$ and $\hat{M}_{\mathrm{eff}}$ validates the extracted parameters.}
  \label{Fig:MUN}
\end{figure}

Application of our methodology to the Ronchi and Sari data yields the intrinsic GB mobility parameters shown in \cref{Fig:MUN}. Global nonlinear regression of the partial-coupling pore drag model (\cref{Eq:Meff_forward}) to the temperature-dependent effective mobility yields:
\begin{equation}
Q_M = 0.87~\text{eV}, \quad M_0 = 1.75 \times 10^{-15}~\text{m}^4/(\text{J}\cdot\text{s}), \quad \phi = 1.00.
\end{equation}
These parameters are nearly identical to those obtained from the effective mobility Arrhenius fit ($Q_{\mathrm{eff}} = 0.89~\text{eV}$, $M_{0,\mathrm{eff}} = 2.05 \times 10^{-15}~\text{m}^4/(\text{J}\cdot\text{s})$), differing by less than 3\% in activation energy and 15\% in prefactor.

The fitted coupling parameter converged to the upper bound $\phi = 1.00$, corresponding to full coupling. Table~\ref{tab:drag_diagnostics_gbmob} shows the self-consistent drag diagnostics computed from the fitted parameters. The dimensionless drag parameter ranges from $\lambda \approx 0.006$ at 1980~K to $\lambda \approx 0.074$ at 1427~K, yielding reduction factors $s \approx 0.93$--$0.99$. This indicates that pore drag reduces GB mobility by at most 7\% relative to the intrinsic value.

\begin{table}[h]
\centering
\footnotesize
\caption{Self-consistent pore drag diagnostics for uranium mononitride grain growth.}
\label{tab:drag_diagnostics_gbmob}
\begin{tabular}{ccccc}
\hline
$T$ (K) & $D$ ($\mu$m) & $r_p$ ($\mu$m) & $\lambda$ & $s$ \\
\hline
1980 & 10.2 & 2.76 & 0.006 & 0.994 \\
1745 & 8.30 & 2.24 & 0.011 & 0.989 \\
1553 & 7.35 & 1.99 & 0.035 & 0.966 \\
1541 & 7.19 & 1.94 & 0.035 & 0.966 \\
1427 & 6.47 & 1.75 & 0.074 & 0.931 \\
\hline
\end{tabular}
\end{table}
Given the order-of-magnitude uncertainties in the surface diffusivity estimate (drawn from UO$_2$ data), the spherical pore approximation, and the limited experimental dataset (five temperatures at a single annealing time), the fitted reduction factors $s \approx 0.93$--$0.99$ are statistically indistinguishable from unity. For this reason, in the following analysis we assume the weak-drag limit and adopt the effective mobility as the intrinsic mobility:
\begin{equation}
M(T) = M_{\mathrm{eff}}(T) = 2.05 \times 10^{-15} \exp\!\left(-\frac{0.89~\text{eV}}{k_B T}\right)~\text{m}^4/(\text{J}\cdot\text{s}).
\label{Eq:M_final_UN}
\end{equation}
This choice is justified by the near-unity reduction factors from the drag model and is further validated by the following exponent analysis.

We tested whether the experimental data can distinguish between cubic ($n=3$) and parabolic ($n=2$) grain growth kinetics. Fitting both models yields $R^2 = 0.970$ (cubic) and $R^2 = 0.961$ (parabolic), with $\Delta R^2 = 0.009 < 0.01$. The statistical equivalence of both kinetic models confirms that the experimental kinetics are indistinguishable from ideal parabolic growth. We therefore conclude that pore drag effects are negligible in the specific samples and microstructural conditions examined by Ronchi and Sari, and the effective mobility measured from their experiments directly represents the intrinsic GB mobility, as given in \cref{Eq:M_final_UN}.

% Had the parabolic model been adopted directly, the extracted mobility parameters would be $M_0 = 3.88 \times 10^{-15}$~m$^4$/(J$\cdot$s), and $Q_M$ = 0.96~eV, which is also close to the $M_\text{eff}$ parameters. However, we retain the $M_{\mathrm{eff}}$ values from the cubic formulation because direct use of parabolic kinetics would require the approximation: $(D^3 - D_0^3)/D(T) = D^2 - D_0^3/D(T) \approx D^2 - D_0^2$ in the Powers and Glaeser relation (\cref{Eq:PG_formulation}), which introduces unnecessary error when $D(T)$ is significantly larger than $D_0$.

\subsection{Phase-field modeling}
\label{sec:PF}

\begin{figure}[h!]
\centering
\begin{subfigure}{0.4\textwidth}
    \includegraphics[width=\linewidth]{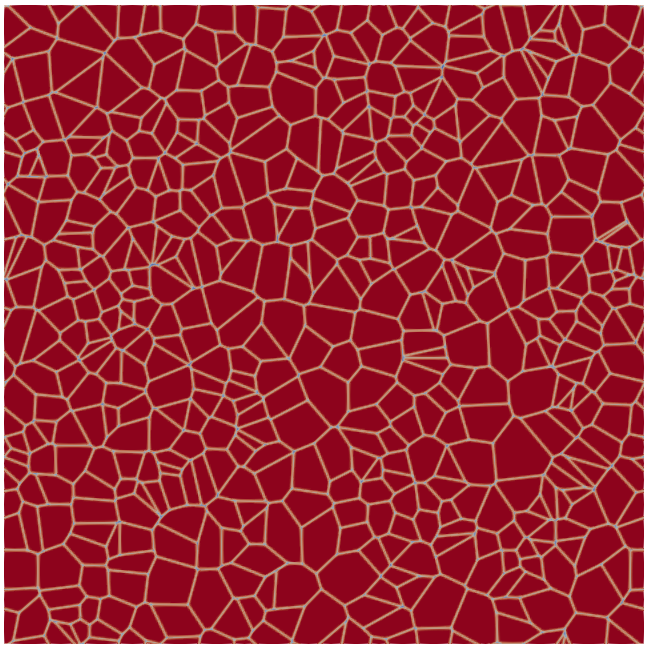}
    \caption{$t$ = 0 s}
    \label{Fig:1500-1}
\end{subfigure}
\hspace{1em}
\begin{subfigure}{0.4\textwidth}
    \includegraphics[width=\linewidth]{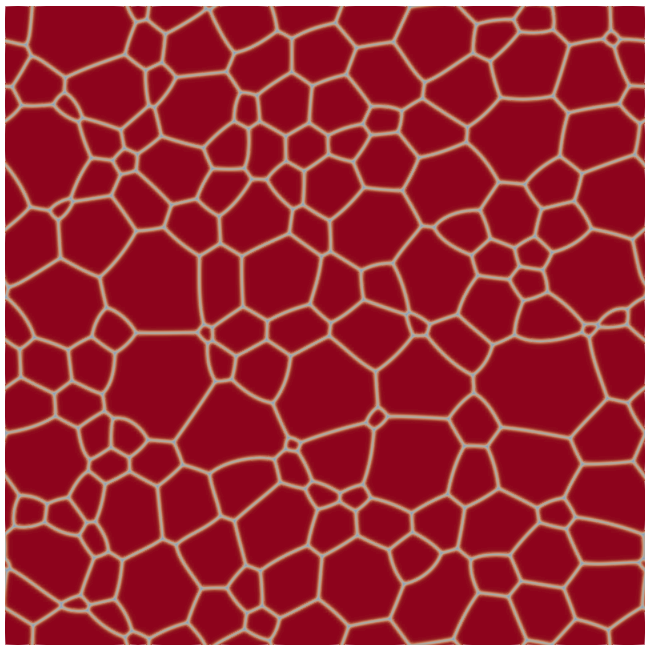}
    \caption{$t$ = $10^4$ s}
    \label{Fig:1500-2}
\end{subfigure}
\caption{Microstructural evolution at 1500~K during grain growth. (\textbf{a}) The initial fine-grained structure at $t = 0$~s and (\textbf{b}) the coarsened structure after $10^4$~s.}
\label{Fig:Grains}
\end{figure}

\begin{figure}[h!]
\centering
\begin{subfigure}{0.48\textwidth}
    \includegraphics[width=\linewidth]{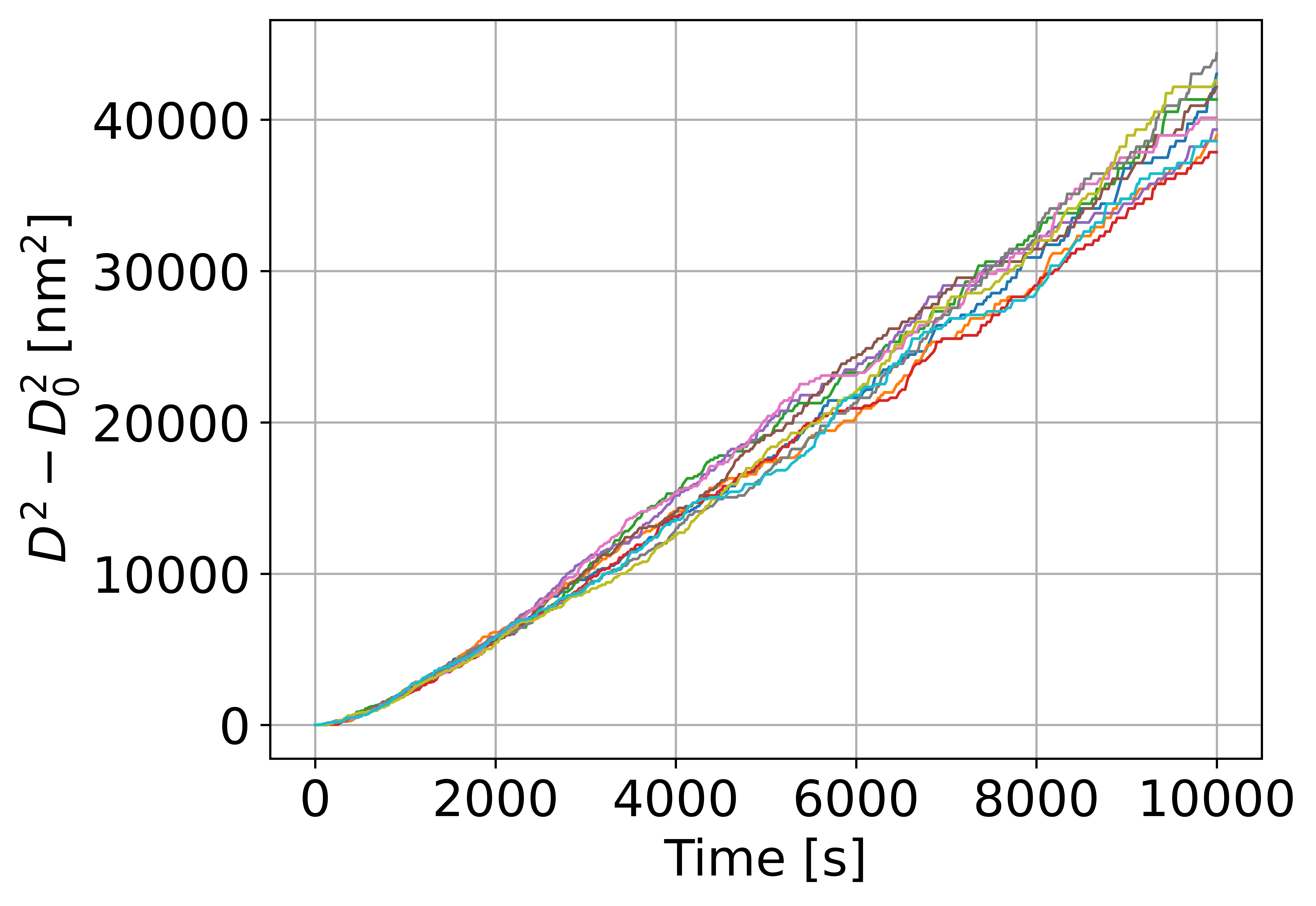}
    \caption{}
    \label{Fig:D2-1500}
\end{subfigure}
\hfill
\begin{subfigure}{0.48\textwidth}
    \includegraphics[width=\linewidth]{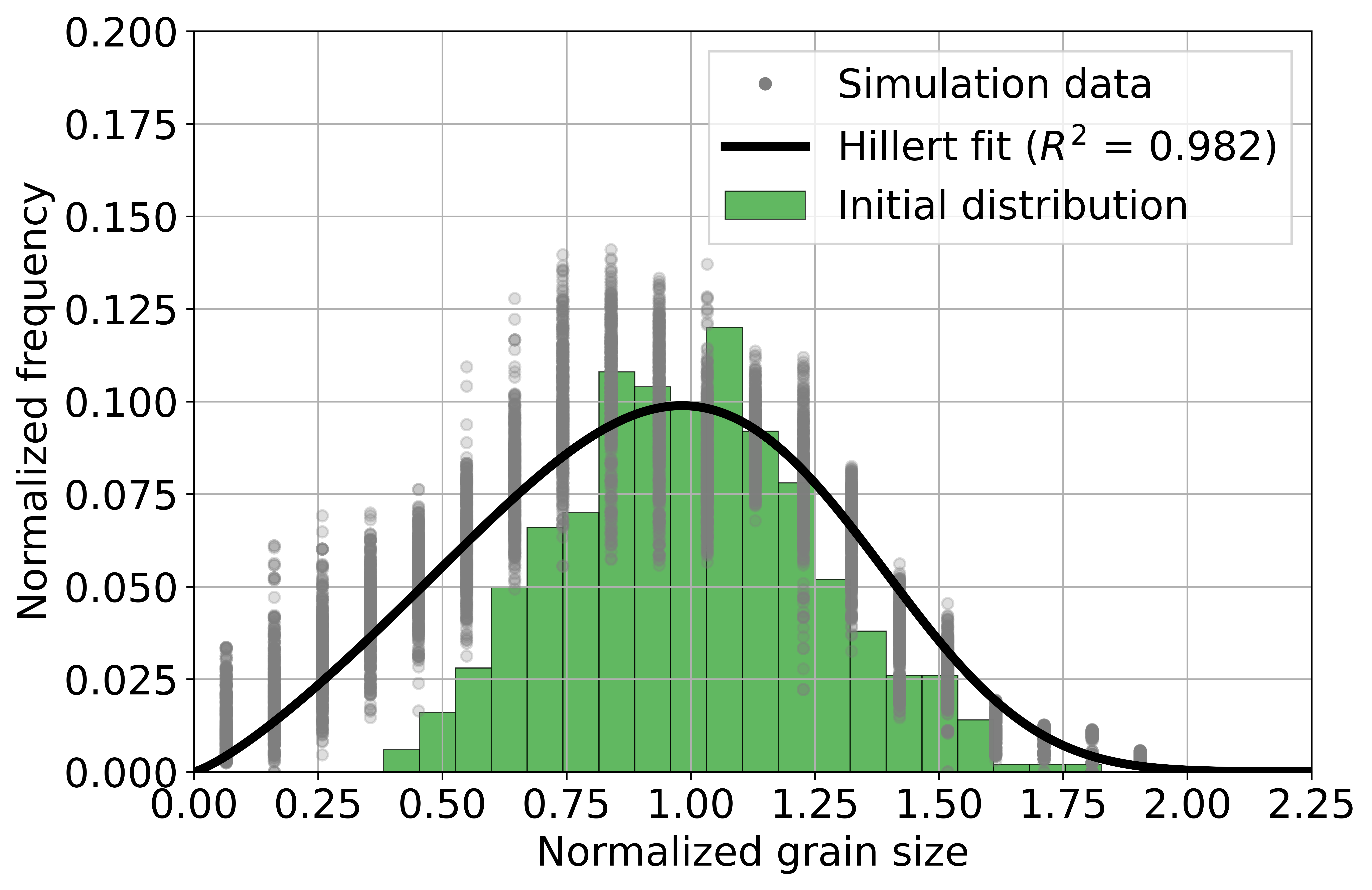}
    \caption{}
    \label{Fig:Hist}
\end{subfigure}
\caption{\textbf{(a)} Simulation results for the time evolution of $(D^2 - D_0^2)$ at 1500~K using 10 different initial microstructures. \textbf{(b)} Grain size evolution at 2000~K between time steps 100--500 (corresponding to about 300--1900 s) represented by normalized grain size distributions (gray points) fitted to a Hillert-like distribution (dark line). The initial grain size distribution (green histogram) is also shown for comparison.}
\end{figure}

Representative grain structures at the beginning of the simulation and after $10^4$~s of growth at 1500~K are shown in \cref{Fig:Grains}. These images clearly illustrate the substantial grain coarsening that occurs during the simulation. The corresponding time evolution of $(D^2 - D_0^2)$ for all initial microstructures at 1500~K is presented in \cref{Fig:D2-1500}. It can be seen that the $(D^2 - D_0^2)$ curves are linear for all initial microstructures, confirming that normal grain growth is observed. To further verify this observation, we plotted the normalized grain size distributions for 2000~K between time steps 100--500 (corresponding to about 300--1900~s) in \cref{Fig:Hist}. As expected, all normalized grain size distributions follow the Hillert-like distribution. Independent of the initial distribution (e.g., the green histogram in \cref{Fig:Hist}), the grain size distribution quickly shifts to the positively skewed Hillert-like distribution typical of normal grain growth. The fitted parameters for the Hillert-like distribution in \cref{Eq:Hillert} at 2000~K averaged over all initial microstructures are shown in \cref{Tab:Hillert}. Similar trends have been observed for all initial microstructures at all temperatures.

\begin{table}[h!]
\centering
\caption{Fitted parameters for the Hillert-like distribution, i.e., \cref{Eq:Hillert}, at 2000~K averaged over all initial microstructures.}
\footnotesize
\begin{tabular}{cc}
\hline 
Parameter & Value \\
\hline
$C$ & $0.1281 \pm 0.0181$ \\
$m$ & $1.3682 \pm 0.2954$ \\
$n$ & $0.3315 \pm 0.1221$ \\
$b$ & $4.0175 \pm 0.5172$ \\
\hline
\end{tabular}
\label{Tab:Hillert}
\end{table}

\begin{figure}[h!]
  \centering
  \includegraphics[width=0.6\textwidth]{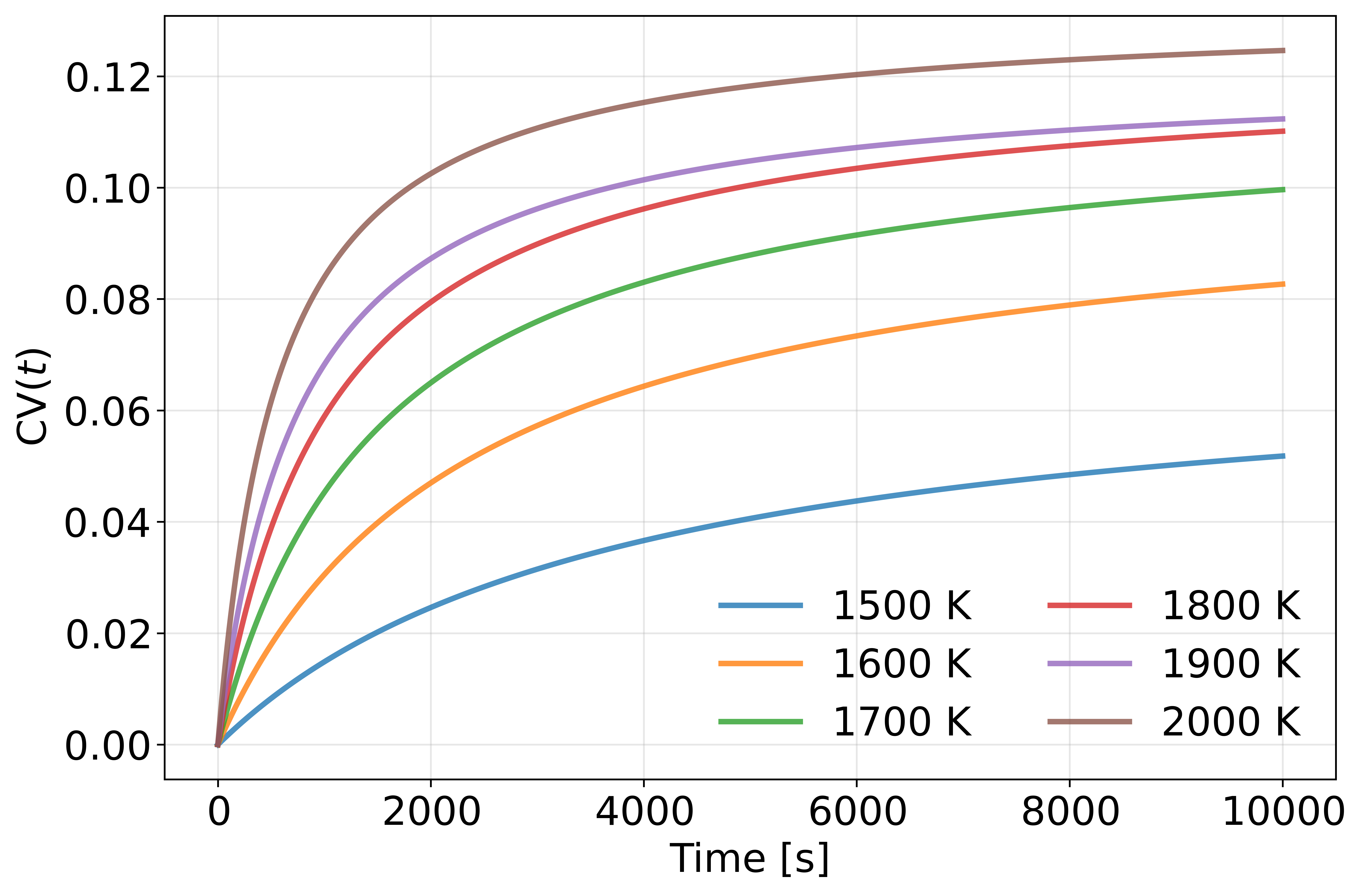}
  \caption{Time evolution of the coefficient of variation at all simulated temperatures.} 
  \label{Fig:CV}
\end{figure}

To quantify the sensitivity of the predicted kinetics to the initial microstructure, we define the coefficient of variation as:
\begin{equation}
\mathrm{CV}(t) = \frac{3\sigma(t)}{\mu(t)},
\end{equation}
where $\sigma(t)$ is the sample standard deviation and $\mu(t)$ is the mean grain size across all ten independent simulations at time $t$. The $3\sigma$ confidence interval normalized by the mean captures approximately 99.7\% of the distribution assuming normality.

\cref{Fig:CV} shows the evolution of the coefficient of variation as a function of time for all temperatures. The CV exhibits a consistent trend across all temperatures: starting from near zero at $t = 0$, CV($t$) increases as grain growth proceeds, reflecting the early-stage divergence among systems with different initial grain size distributions. As coarsening continues, CV($t$) gradually approaches an asymptotic plateau, signaling the achievement of self-similarity in the grain-size distribution.

To quantify this behavior, we fit the CV evolution to an exponential saturation model \cite{AbdulHameed2024c}:
\begin{equation}
\mathrm{CV}(t) = \mathrm{CV}_0 \left(1 - e^{-\beta t}\right),
\end{equation}
where $\mathrm{CV}_0$ is the asymptotic coefficient of variation and $\beta$ is the rate constant characterizing the approach to steady state. The fitted parameters reveal a systematic temperature dependence, summarized in \cref{Tab:CV_params}. The asymptotic CV increases monotonically with temperature, from $\mathrm{CV}_0 \approx 0.053$ at 1500~K to $\mathrm{CV}_0 \approx 0.120$ at 2000~K, indicating that higher temperatures produce broader grain size distributions at steady state. Simultaneously, the rate constant $\beta$ increases from $2.98 \times 10^{-4}$~s$^{-1}$ at 1500~K to $1.14 \times 10^{-3}$~s$^{-1}$ at 2000~K, demonstrating that higher temperatures accelerate the approach to self-similarity. The characteristic time to reach steady state, $\tau = 1/\beta$, decreases from approximately 3360~s at 1500~K to 875~s at 2000~K.

This rise and subsequent saturation of CV has been reported in previous studies, such as the work by Breithaupt \textit{et al.}~\cite{Breithaupt2021}, who found that the CV saturates at approximately $0.40$ for all initial conditions considered. While our CV quantifies variability across simulations rather than within a single microstructure as in Breithaupt \textit{et al.}, the qualitative behavior, i.e., monotonic increase followed by exponential saturation, demonstrates that both inter-simulation and intra-microstructure heterogeneity exhibit analogous evolution toward self-similarity.

In summary, this analysis demonstrates that while the influence of the initial grain size distribution persists, it becomes bounded at long times as the system approaches self-similarity. The temperature-dependent asymptotic CV values and rate constants reveal that higher temperatures promote both greater steady-state heterogeneity and faster kinetics toward the self-similar regime.

\begin{table}[htbp]
\footnotesize
\centering
\caption{Fitted parameters for the coefficient of variation evolution.}
\label{Tab:CV_params}
\begin{tabular}{ccc}
\hline
Temperature (K) & $\mathrm{CV}_0$ & $\beta$ (s$^{-1}$) \\
\hline
1500 & 0.0533 & $2.98 \times 10^{-4}$ \\
1600 & 0.0814 & $4.15 \times 10^{-4}$ \\
1700 & 0.0963 & $5.53 \times 10^{-4}$ \\
1800 & 0.1059 & $7.14 \times 10^{-4}$ \\
1900 & 0.1079 & $9.04 \times 10^{-4}$ \\
2000 & 0.1199 & $1.14 \times 10^{-3}$ \\
\hline
\end{tabular}
\end{table}

\subsection{Uncertainty quantification and sensitivity analysis}\label{sec:UQ}

\subsubsection*{Grain size distribution and analysis of extremes}

\begin{figure}[h!]
\centering
\begin{subfigure}{0.48\textwidth}
    \includegraphics[width=\linewidth]{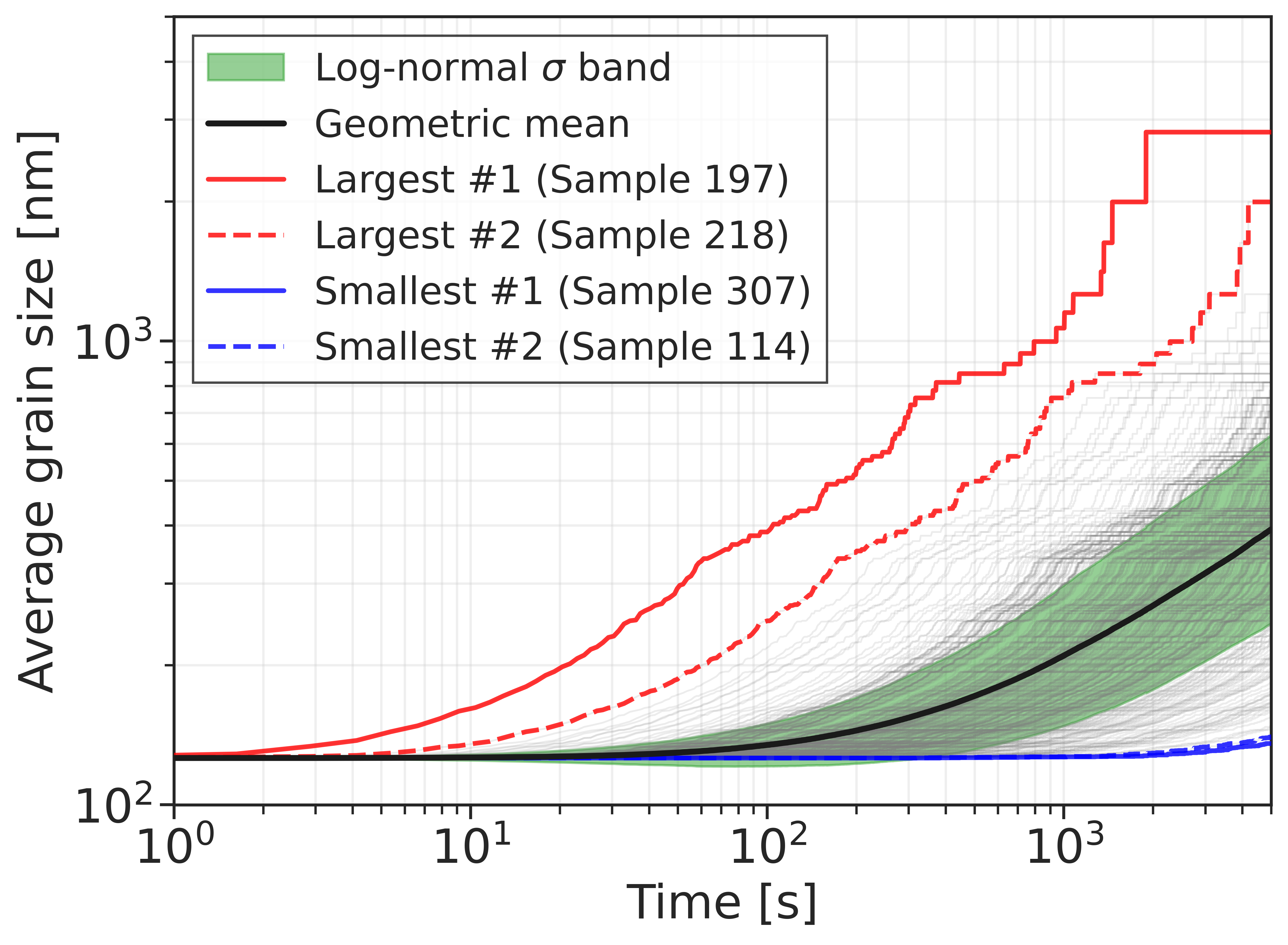}
    \caption{}
    \label{Fig:Trajectories}
\end{subfigure}
\hfill
\begin{subfigure}{0.48\textwidth}
    \includegraphics[width=\linewidth]{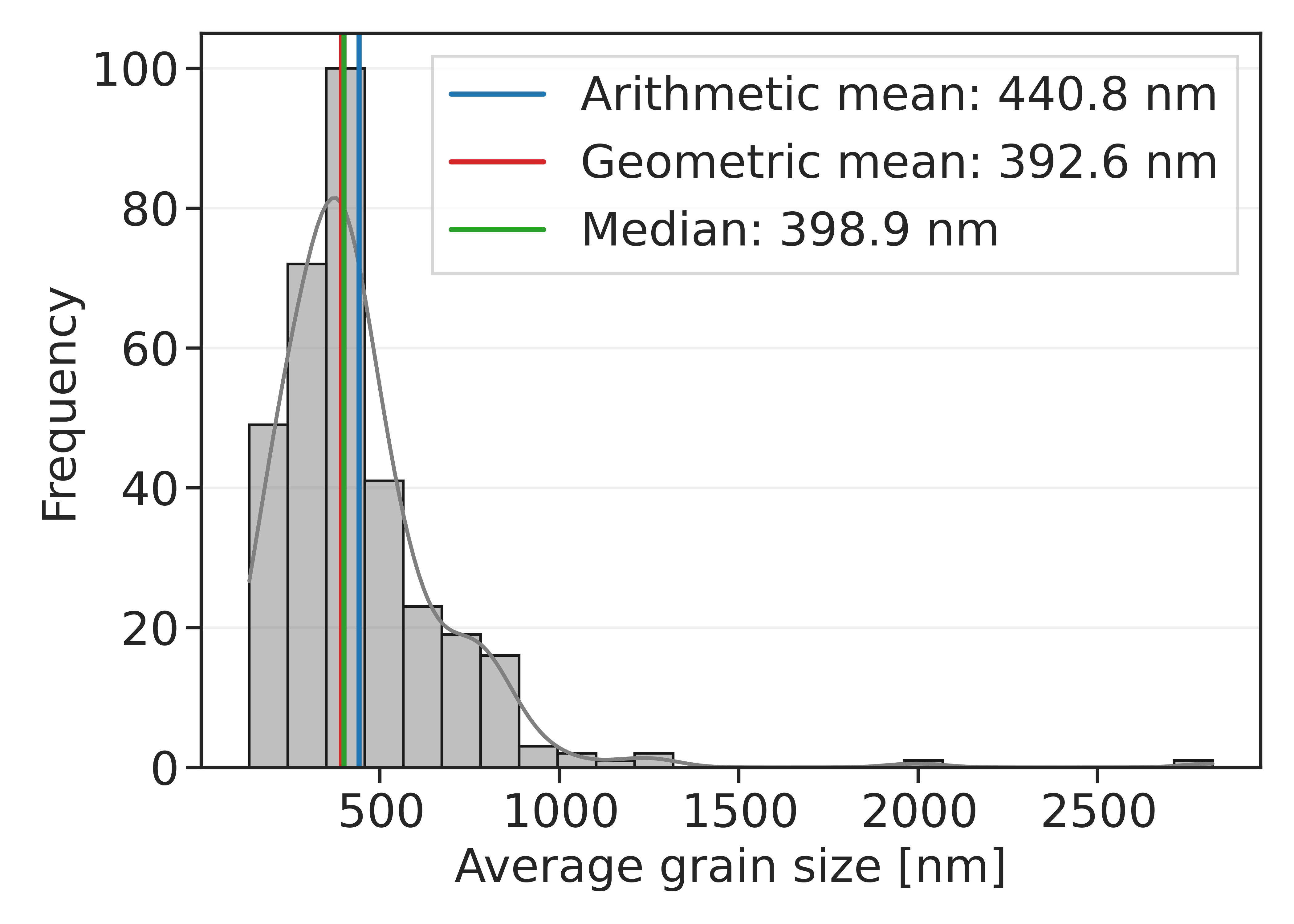}
    \caption{}
    \label{Fig:d_dist}
\end{subfigure}
\caption{\textbf{(a)} Grain growth trajectories: gray lines show all 330 simulations, the shaded band represents the log-normal $\sigma$ interval, and the black line shows the geometric mean. Red lines indicate the two largest final grain sizes, while blue lines indicate the two smallest. \textit{(b)} Distribution of final grain sizes at $t = 5000$ s. The histogram shows right skewness characteristic of log-normal distributions.}
\end{figure}

As shown in \cref{Fig:d_dist}, the final grain size distribution at $t = 5000$ s is approximately log-normal. To understand why, note that:
\begin{equation}
\ln D = \text{const} + \frac{1}{2}\ln M_0 + \frac{1}{2}\ln \gamma - \frac{Q_M}{2k_B T}.
\end{equation}
The dominant contributions to $\ln D$ are Gaussian: $\ln M_0$ is normally distributed by construction, and $Q_M$ follows a weakly truncated normal distribution. The remaining term, $\frac{1}{2}\ln \gamma$, is assessed using the delta method~\cite{Dorfman1938, Cramer1946}. We expand $\ln \gamma$ about $\mu_\gamma$,
\begin{equation}
\frac{1}{2} \ln \gamma = \frac{1}{2} \left( \ln \mu_\gamma + \frac{\gamma - \mu_\gamma}{\mu_\gamma} - \frac{(\gamma - \mu_\gamma)^2}{2\mu_\gamma^2} + \cdots \right)
\end{equation}
The linear term is a normal random variable. The quadratic term has expected magnitude scaling as $\sigma_\gamma^2 / \mu_\gamma^2 \approx 0.088$, and after the prefactor of $1/2$ in $\ln D$, its net contribution is approximately $4$--$5\%$. The non-Gaussian distortion from $\ln \gamma$ is therefore small relative to the dominant Gaussian terms from $\ln M_0$ and $Q_M$, so $\ln D$ is approximately normal and $D$ is approximately log-normal, consistent with the right-skewed histogram observed in \cref{Fig:d_dist}.

\begin{figure}[h!]
\centering
\begin{subfigure}{0.48\textwidth}
    \includegraphics[width=\linewidth]{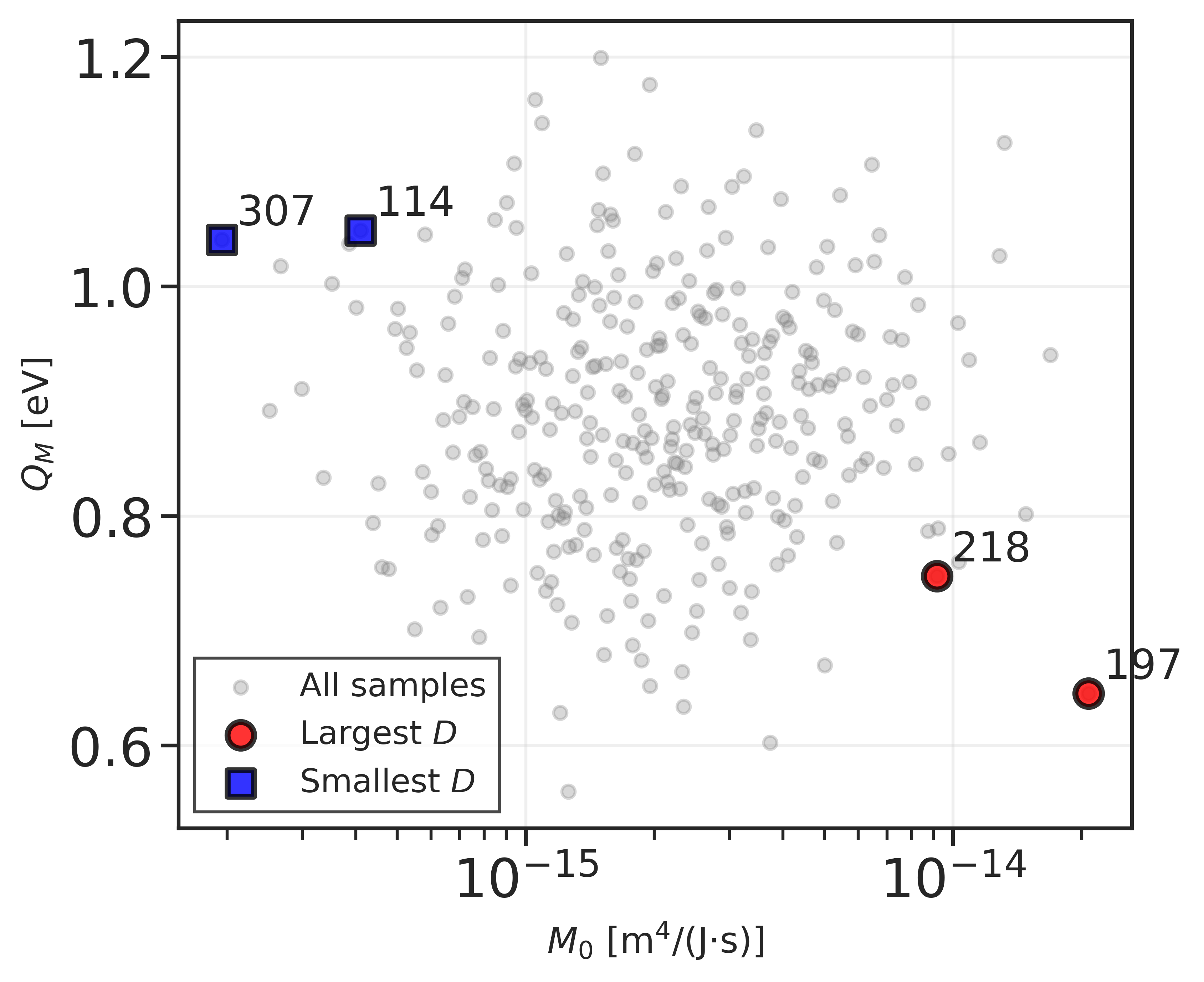}
    \caption{$M_0$ vs. $Q_M$}
    \label{Fig:M0vsQM}
\end{subfigure}
\hfill
\begin{subfigure}{0.48\textwidth}
    \includegraphics[width=\linewidth]{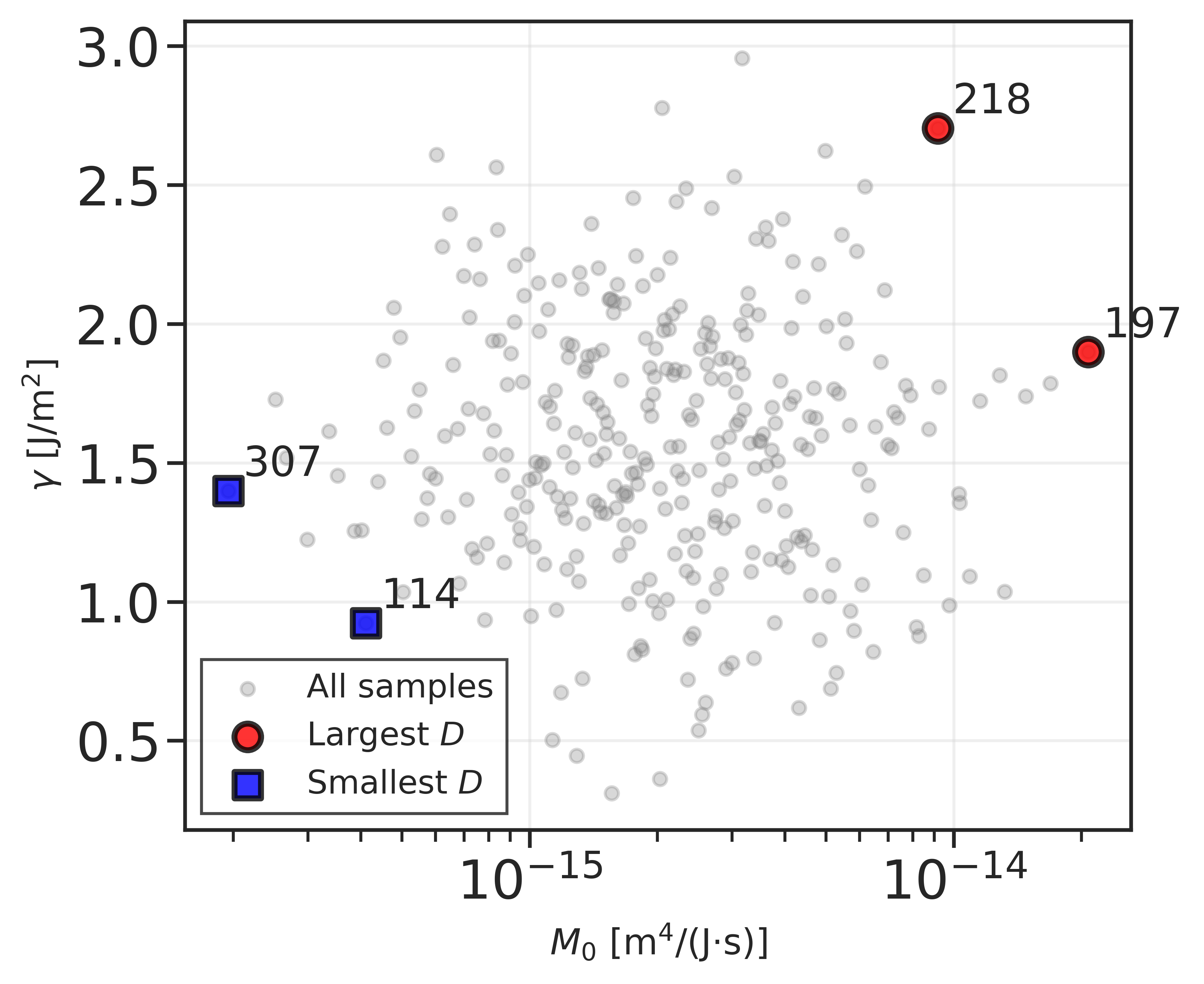}
    \caption{$M_0$ vs. $\gamma$}
    \label{Fig:M0vsgamma}
\end{subfigure}
\hfill
\begin{subfigure}{0.48\textwidth}
    \includegraphics[width=\linewidth]{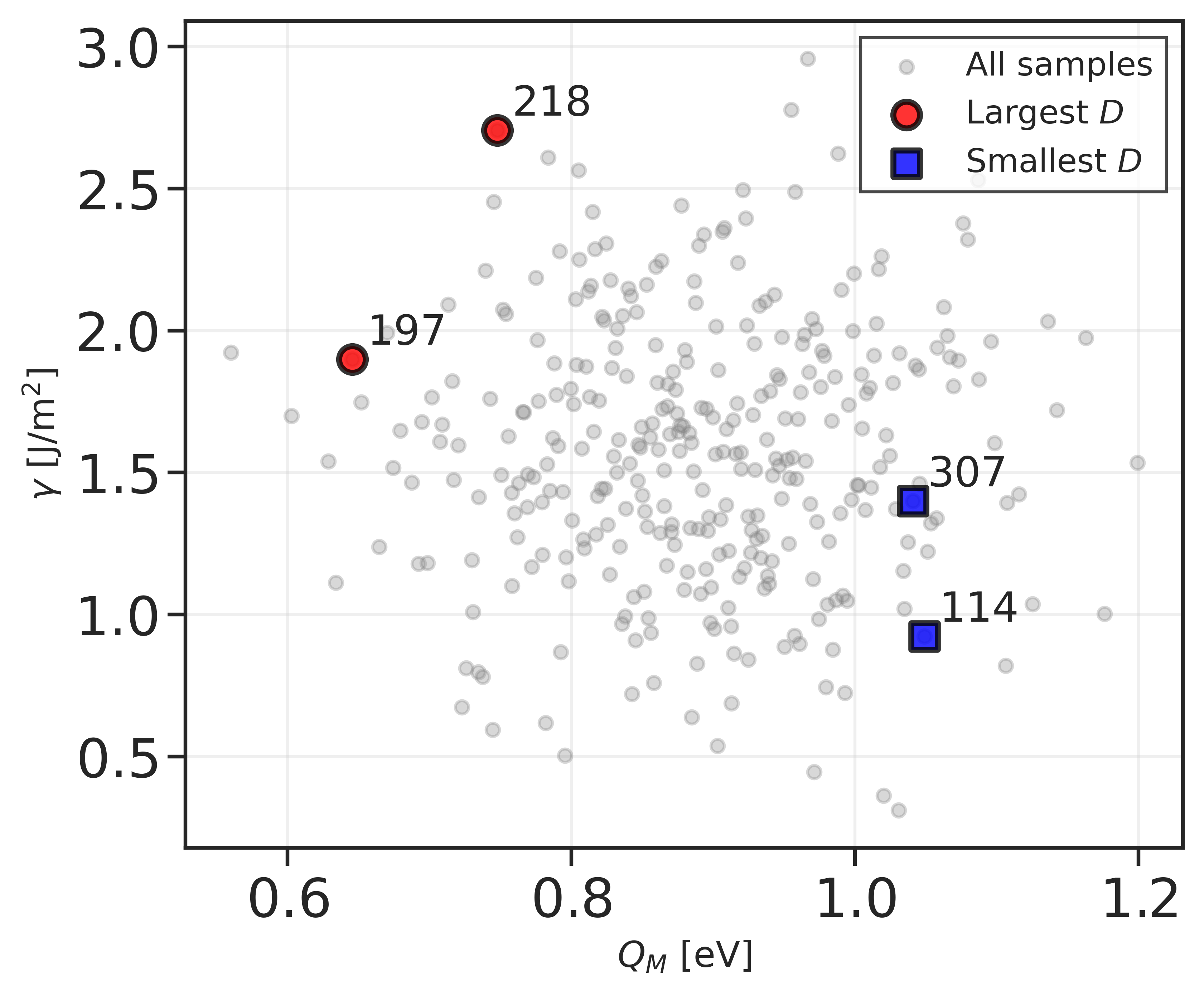}
    \caption{$Q_M$ vs. $\gamma$}
    \label{Fig:QMvsgamma}
\end{subfigure}
\caption{Parameter space locations of extreme grain sizes.}
\end{figure}

The grain size distribution exhibits two orders of magnitude spread (\cref{Fig:d_dist,Fig:Trajectories}). As shown in \cref{Fig:M0vsQM}, the two largest final grain sizes (2821 nm and 1995 nm) result from high $M_0$ and low $Q_M$: sample 197 has $M_0 = 2.08 \times 10^{-14}$ m$^4$/(J$\cdot$s) and $Q_M = 0.645$ eV, while sample 218 has $M_0 = 9.17 \times 10^{-15}$ m$^4$/(J$\cdot$s) and $Q_M = 0.748$ eV. This combination yields high effective mobility, producing rapid grain growth that depletes the grain population early (\cref{Fig:Trajectories}). Conversely, the smallest grain sizes (136 nm and 140 nm) arise from low $M_0$ and high $Q_M$: sample 307 has $M_0 = 1.95 \times 10^{-16}$ m$^4$/(J$\cdot$s) and $Q_M = 1.04$ eV, while sample 114 has $M_0 = 4.11 \times 10^{-16}$ m$^4$/(J$\cdot$s) and $Q_M = 1.05$ eV (1.18$\times$ mean). As shown in \cref{Fig:M0vsgamma,Fig:QMvsgamma}, the GB energy $\gamma$ varies between 0.59 and 1.72 times the mean across extreme cases without consistently correlating to grain size, confirming its weaker influence relative to $M_0$ and $Q_M$ identified in detail later in the Sobol analysis.

\subsubsection*{PCA dimensionality reduction}

PCA of the grain size evolution trajectories across 330 simulation runs and up to 5000~s reveals that the variance is highly concentrated in the first few components. PC1 explains 95.94\% of the total variance, while PC2 explains 2.98\% of the total variance. Together, these PCs retain 98.92\% of the original signal variance. The reconstruction diagnostics show a relative root mean square error (RMSE) of nearly 0 when using the first two PCs to reconstruct the full trajectories, indicating that they capture nearly all relevant dynamical features of grain growth, with PC1 describing the dominant growth trend and PC2 capturing secondary deviations from this trend.

\subsubsection*{GP surrogate model performance}

Gaussian process surrogate models were trained to map the input parameter space ($M_0$, $Q_M$, $\gamma$) to the PC scores. The predictive capability of these models was rigorously assessed using a 25-fold repeated five-fold cross-validation strategy. The models demonstrate exceptional accuracy across all splits. For PC1, the minimum observed $R^2$ across all splits was 99.996\%, and the minimum $Q^2$ was 98.46\%, indicating near-perfect reconstruction of the dominant growth mode. For PC2, which captures finer, lower-magnitude variations, the minimum $R^2$ was 99.972\% and the minimum $Q^2$ was 90.34\%, reflecting somewhat greater difficulty in predicting the secondary mode but still maintaining strong generalization capability. The learned noise variances in standardized PC space are $8.74 \times 10^{-5}$ for PC1 and $5.26 \times 10^{-4}$ for PC2, confirming that the surrogate models successfully learn smooth, low-noise representations of the underlying phase-field simulations.

\subsubsection*{Global sensitivity analysis}

\begin{figure}[h!]
\centering
\begin{subfigure}{0.48\textwidth}
    \includegraphics[width=\linewidth]{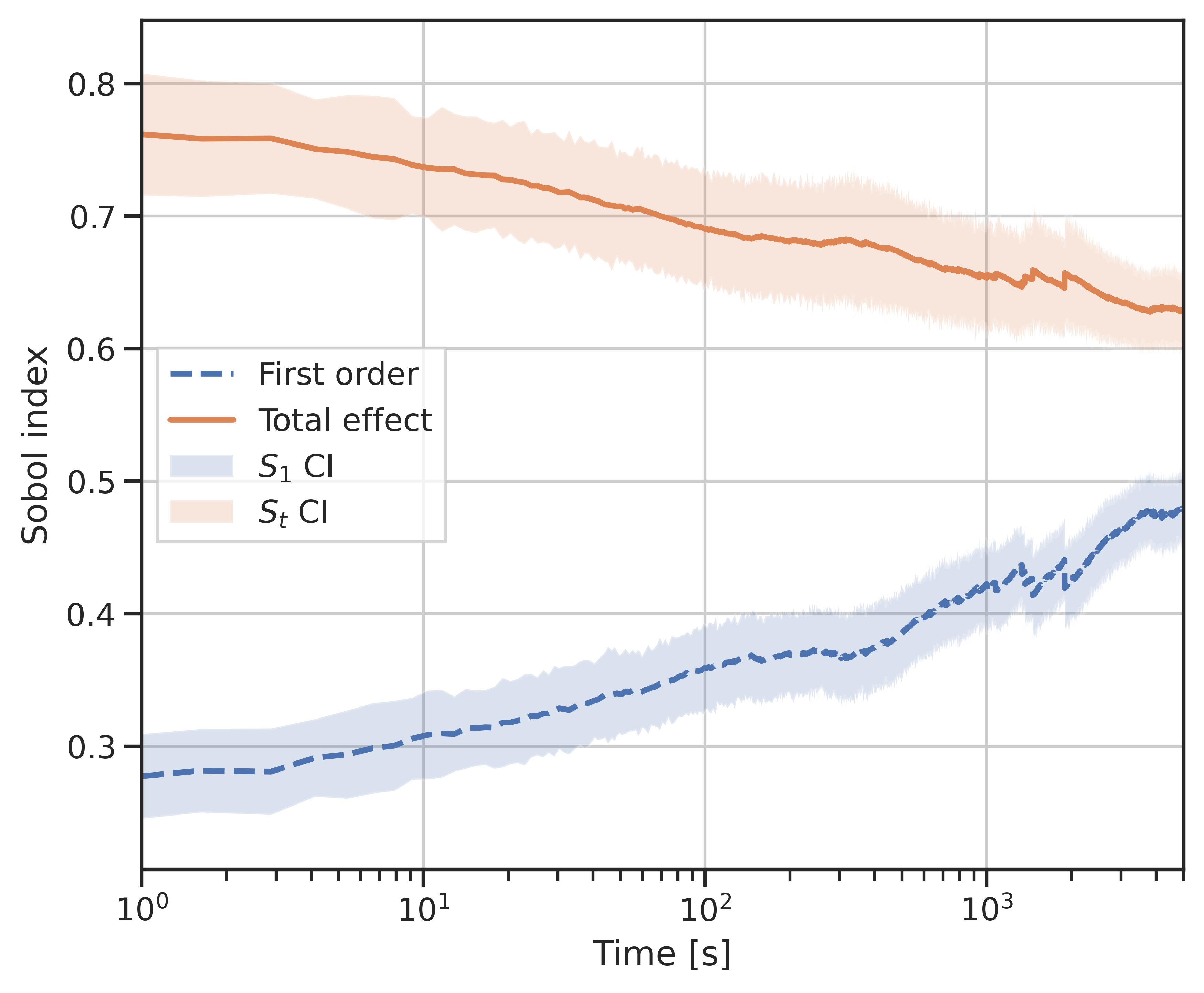}
    \caption{$M_0$}
\end{subfigure}
\hfill
\begin{subfigure}{0.48\textwidth}
    \includegraphics[width=\linewidth]{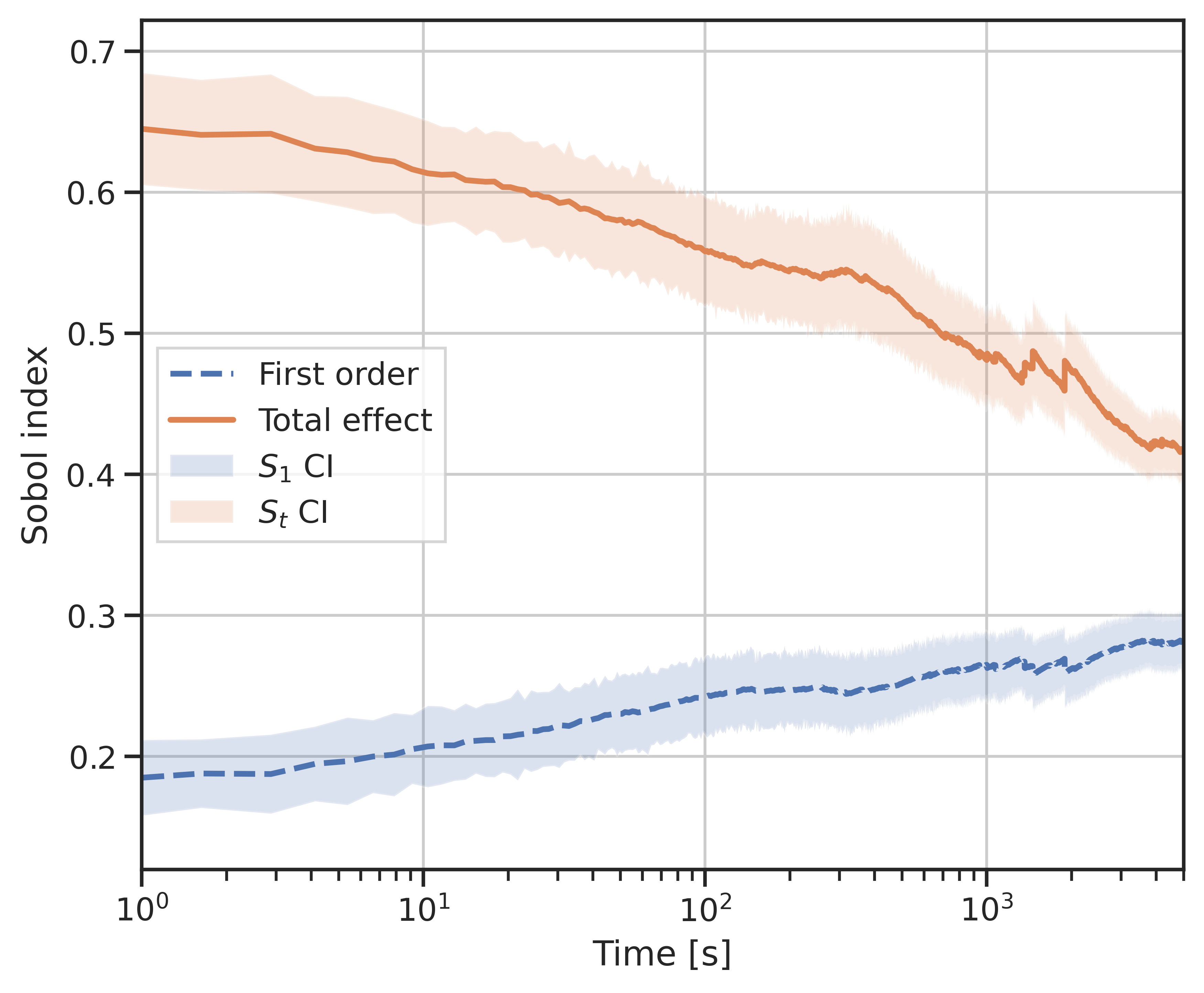}
    \caption{$Q_M$}
\end{subfigure}

\begin{subfigure}{0.48\textwidth}
    \includegraphics[width=\linewidth]{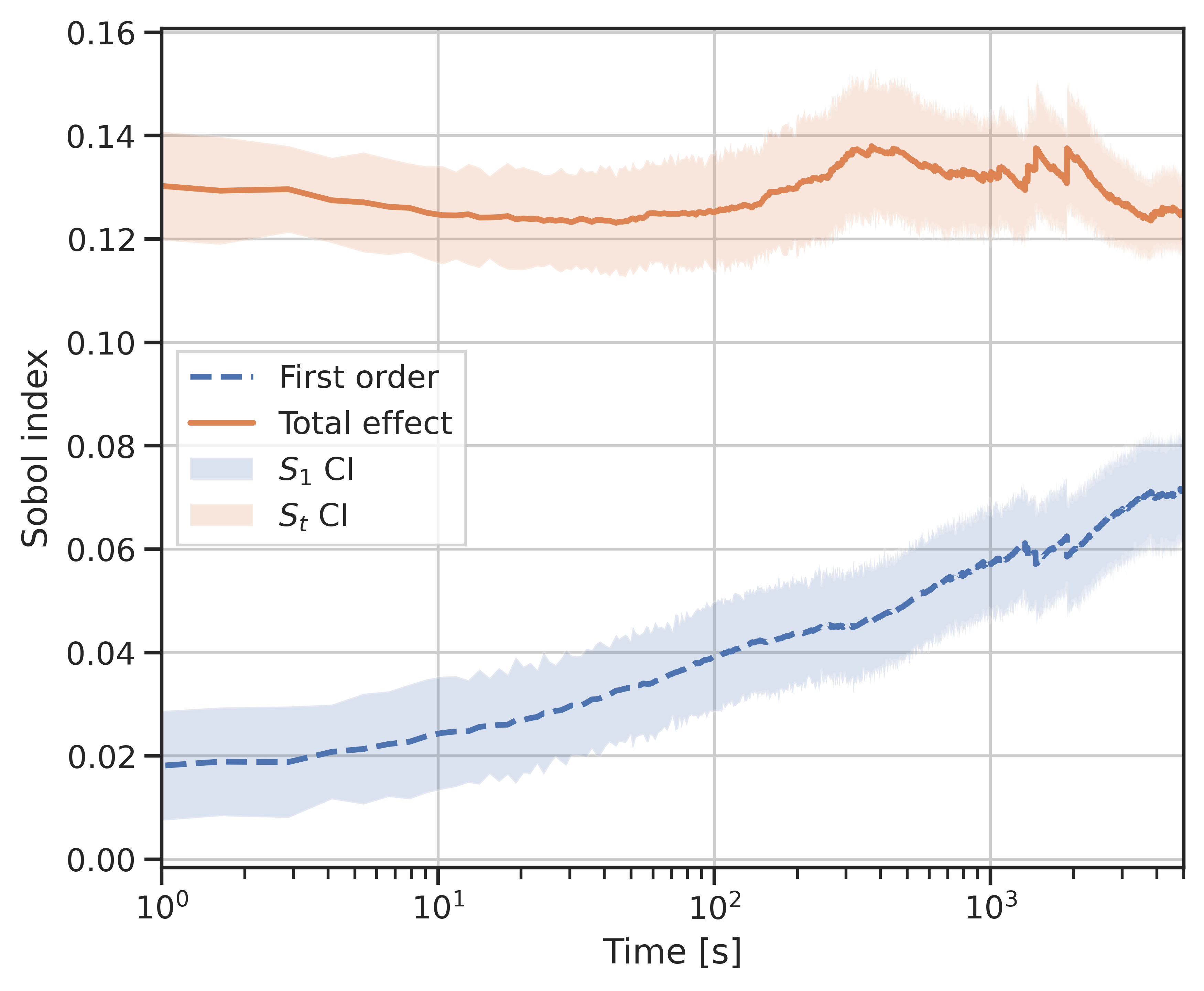}
    \caption{$\gamma$}
\end{subfigure}
\caption{Time-resolved Sobol sensitivity indices for: \textbf{(a)} mobility prefactor $M_0$, \textbf{(b)} activation energy $Q_M$, and \textbf{(c)} GB energy $\gamma$. Solid lines represent total-effect indices ($S_t$), dashed lines represent first-order indices ($S_1$), and shaded regions indicate 95\% confidence intervals (CI) from \texttt|SALib|.}
\label{Fig:Sobol}
\end{figure}

Time-resolved Sobol indices reveal distinct temporal evolution patterns for the three uncertain input parameters (\cref{Fig:Sobol}). At the earliest time point ($t \approx 0$~s), $M_0$ exhibits the highest first-order sensitivity index ($S_1 = 0.268$) and total effect ($S_t = 0.769$), followed by $Q_M$ ($S_1 = 0.179$, $S_t = 0.653$) and $\gamma$ ($S_1 = 0.017$, $S_t = 0.132$). The large gap between $S_1$ and $S_t$ for all three parameters indicates substantial higher-order interactions at early times, when the system has not yet reached the asymptotic parabolic growth regime.

As grain growth progresses, the sensitivity indices evolve systematically. The first-order sensitivity of $M_0$ increases monotonically from $S_1 = 0.268$ at $t \approx 0$~s to $S_1 = 0.481$ at $t = 5000$~s, while its total effect decreases from $S_t = 0.769$ to $S_t = 0.629$. Conversely, $Q_M$ exhibits an initial increase in $S_1$ from 0.179 to approximately 0.282 by $t \approx 1300$~s, followed by a plateau that stabilizes near $S_1 \approx 0.28$. For $\gamma$, the first-order sensitivity increases from $S_1 = 0.017$ at early times to $S_1 = 0.072$ at late times, while its total effect remains nearly constant at $S_t \approx 0.12$--$0.13$ throughout the entire simulation.

The convergence of $S_1$ and $S_t$ for all three parameters at late times indicates diminishing higher-order interactions as the system approaches the asymptotic parabolic regime described by \cref{Eq:ideal}. At $t = 5000$~s, the sum of first-order indices approaches unity ($\sum S_1 \approx 0.83$), confirming that parameter interactions become negligible compared to individual contributions. However, a notable gap persists between the sum of first-order indices and unity, indicating that approximately 15--20\% of the system variance is driven by nonlinear interactions between the parameters rather than their isolated effects. For instance, at $t = 4038$~s, the total effect for $M_0$ is $S_t = 0.631$ while its first-order index is $S_1 = 0.476$, demonstrating that interactions contribute roughly $S_t - S_1 = 0.155$ (or 15.5\%) to the total sensitivity of $M_0$.

The sensitivity hierarchy $S_1(M_0) > S_1(Q_M) > S_1(\gamma)$ is primarily driven by the Arrhenius dependence $M = M_0 \exp(-Q_M / (k_B T))$, through which uncertainties in $M_0$ and $Q_M$ are exponentially amplified, whereas $\gamma$ enters the growth rate linearly and with comparatively smaller absolute uncertainty. At late times ($t = 5000$~s), $M_0$ accounts for approximately 48\% of output variance in first-order effects and 63\% including interactions, $Q_M$ contributes 28\% (first-order) to 42\% (total), and $\gamma$ accounts for only 7\% (first-order) to 12\% (total). These findings highlight the critical importance of accurate mobility measurements for predictive grain growth modeling in UN, and suggest that future experimental efforts should prioritize reducing uncertainty in $M_0$ and $Q_M$ over further refinement of $\gamma$.

\section{Discussion}

The central methodological contribution of this work is the extraction of a drag-free GB mobility for UN from the only available grain growth dataset~\cite{Ronchi1975}. We show that pore drag is negligible under the microstructural conditions of Ronchi and Sari: the pore-drag analysis yields mobility reduction factors $s \approx 0.93$--$0.99$, and the cubic and parabolic kinetic models are statistically indistinguishable ($\Delta R^2 = 0.009$). The effective mobility therefore directly represents the intrinsic GB mobility.

Three sources of uncertainty limit the precision of this result. First, the surface diffusivity in the pore mobility model (\cref{Eq:Mp_pore}) was approximated from UO$_2$ data~\cite{Muntaha2024}, as no measurements or atomistic calculations exist for UN. Second, the experimental dataset comprises only five temperature points at a single annealing time (500~h), which is insufficient to determine $M_0$ and $Q_M$ with high confidence. Third, the material studied was (U,Pu)N rather than pure UN. Solute drag from Pu cannot be quantified without dedicated experiments. The extracted mobility should therefore be treated as an order-of-magnitude estimate. Dedicated isothermal grain growth experiments on high-purity UN pellets over a wider temperature range are needed to refine these parameters.

\begin{figure}[h!]
  \centering
  \includegraphics[width=0.6\textwidth]{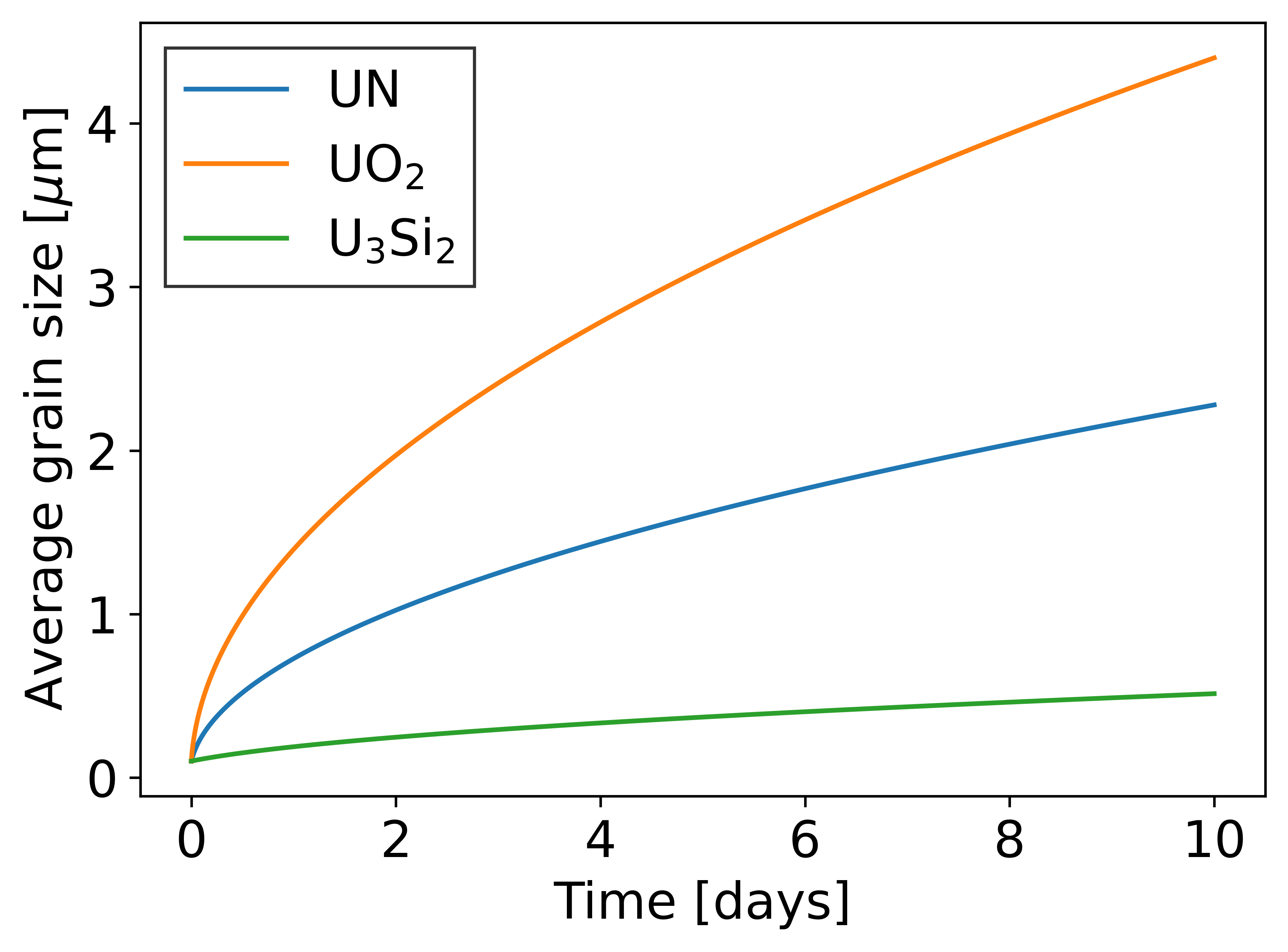}
  \caption{Evolution of average grain size $D(t)$ at $T = 1500$~K for UO$_2$, UN, and U$_3$Si$_2$, assuming drag-free curvature-driven growth with an initial grain size of $D_0 = 0.1~\mu$m.}
  \label{Fig:UN-UO2-U3Si2}
\end{figure}

It is instructive to compare the intrinsic GB mobility of UN with those reported for UO$_2$~\cite{Tonks2014, Bai2015} and U$_3$Si$_2$~\cite{Cheniour2020}, the two other actinide fuel systems for which phase-field grain growth models exist. \cref{Fig:UN-UO2-U3Si2} shows the drag-free grain growth predicted by \cref{Eq:ideal} at 1500~K for all three materials over 10 days. Despite having the largest activation energy by far ($Q_M = 3.01$~eV versus 0.89~eV for UN and 0.33~eV for U$_3$Si$_2$), UO$_2$ grows the fastest, reaching $D \approx 4.40~\mu$m after 10 days. This is a direct consequence of its exceptionally large prefactor ($M_0 = 2.14 \times 10^{-7}$~m$^4$/(J$\cdot$s)), which overwhelms the Arrhenius suppression at this temperature. UN reaches $D \approx 2.28~\mu$m, and U$_3$Si$_2$ grows the least, reaching only $D \approx 0.51~\mu$m. UN occupies an intermediate position between the highly ionic UO$_2$ and the intermetallic U$_3$Si$_2$, reflecting its mixed ionic-covalent bonding character.

The phase-field simulations represent the drag-free limit of grain growth in UN. Solute drag, pore-boundary interactions, irradiation-induced defects, and thermal gradient driving forces are all absent. The predicted grain sizes are therefore upper bounds on curvature-driven coarsening at each temperature. Including these resistive mechanisms is a necessary step toward more realistic simulations.

A limitation of the present simulations is their 2D nature. While 2D grain growth captures the correct qualitative behavior, the geometric factor $\alpha = 1/2$ in \cref{Eq:Curve} differs from the value $\alpha = 1$ appropriate for 3D equiaxed grains, so the predicted growth rates are a factor of two lower than those expected in bulk pellets. Despite this, the 2D framework is sufficient for the objectives of this study---i.e, demonstrating normal grain growth, validating the Hillert-like grain size distribution, and performing the sensitivity analysis---all of which are qualitatively insensitive to dimensionality.

The average GB energy correlation (\cref{Eq:gamma}) fits the MD data well over 0--2000~K ($R^2 = 0.983$), but the $T^5$ term grows rapidly and the fit should not be extrapolated beyond this range.

The initial grain size distribution is not explicitly included in the uncertainty analysis. Instead, its effect is approximated by averaging grain growth over ten independent Voronoi tessellations. A more complete framework would parameterize this distribution explicitly---for instance, through moments of kernel density estimates fitted to each tessellation---and include these as additional uncertain inputs. This would quantify the joint effect of kinetic uncertainty and structural randomness on grain evolution, and represents a natural direction for future work.

Despite the limitations discussed above, this study establishes the first integrated multiscale modeling framework for grain growth in UN. It provides the community with temperature-dependent GB energies for 27 symmetric tilt boundaries, a rigorous and transferable procedure for extracting drag-free GB mobility from scarce experimental data, the first quantitative grain growth kinetics for UN, and clear experimental priorities identified through time-resolved global sensitivity analysis. This physics-informed, uncertainty-aware framework is directly applicable to fuel performance codes and extensible to other advanced nuclear fuel systems.

\section{Conclusions}

This work presents an integrated multiscale framework for grain growth in uranium mononitride, combining MD-based GB energy calculations, phenomenological mobility extraction, phase-field simulations, and surrogate-assisted uncertainty quantification.

The Tseplyaev potential \cite{Tseplyaev2016} yields GB energies consistent with DFT values across all 27 symmetric tilt boundaries studied, whereas the Kocevski potential \cite{Kocevski2022II} systematically underestimates them due to anomalous U-U repulsion that prevents GB closure upon 0~K minimization. The average GB energy is nearly temperature-independent below 1000~K and increases at higher temperatures, well described by:
\begin{equation*}
    \gamma(T) = 1.382 + 3.279\times10^{-6}\,T + 5.748\times10^{-18}\,T^5 \quad (\text{J/m}^2),
\end{equation*}
with $R^2 = 0.983$. The standard deviation of the 27 GB energies is approximately 30\% of the mean at all temperatures, reflecting the spread of GB energies across misorientation angles.

A mechanistic pore-drag analysis, based on the Powers-Glaeser \cite{Powers1998} framework with partial pore-GB coupling, applied to the Ronchi and Sari \cite{Ronchi1975} dataset yields mobility reduction factors $s \approx 0.93$--$0.99$, statistically indistinguishable from unity. The cubic and parabolic kinetic models are equally consistent with the experimental data ($\Delta R^2 = 0.009$), confirming that pore drag is negligible under the microstructural conditions of those experiments. The intrinsic GB mobility is therefore identified with the effective mobility:
\begin{equation*}
    M(T) = 2.05\times10^{-15}\exp\!\left(-\frac{0.89~\text{eV}}{k_B T}\right) \quad \text{m}^4/(\text{J}\cdot\text{s}).
\end{equation*}

Phase-field simulations over 1500--2000~K confirm normal curvature-driven grain growth, with $(D^2 - D_0^2)$ evolving linearly in time and grain size distributions converging to the Hillert-like form independently of the initial microstructure. The coefficient of variation across ten independent Voronoi microstructures saturates exponentially at each temperature, with the asymptotic value increasing from $\mathrm{CV}_0 \approx 0.053$ at 1500~K to $\mathrm{CV}_0 \approx 0.120$ at 2000~K, and the characteristic saturation time decreasing from approximately 3360~s to 875~s over the same range.

The surrogate-assisted global sensitivity analysis, combining PCA, Gaussian process regression, and Sobol decomposition, reveals a clear and persistent sensitivity hierarchy. The mobility prefactor $M_0$ dominates output variance at all times, reaching $S_1 \approx 0.48$ and $S_t \approx 0.63$ at $t = 5000$~s, followed by the activation energy $Q_M$ ($S_1 \approx 0.28$, $S_t \approx 0.42$), while the GB energy $\gamma$ contributes minimally ($S_1 \approx 0.07$, $S_t \approx 0.12$). This hierarchy reflects the exponential amplification of Arrhenius parameter uncertainty relative to the linear entry of $\gamma$ into the grain growth rate. The convergence of $S_1$ and $S_t$ at late times confirms that parameter interactions diminish as the system approaches the asymptotic parabolic growth regime. Reducing uncertainty in $M_0$ and $Q_M$ through dedicated isothermal experiments on high-purity UN over a wide temperature range is therefore the highest-priority need for predictive grain growth modeling in this material.

\section{Acknowledgments}

The authors thank Xu Wu for the fruitful discussions and useful suggestions. This research made use of the resources of the High-Performance Computing Center at Idaho National Laboratory, which is supported by the Office of Nuclear Energy of the U.S. Department of Energy and the Nuclear Science User Facilities under Contract No. DE-AC07-05ID14517.

\FloatBarrier
\bibliographystyle{elsarticle-num}
\bibliography{ref}

\end{document}